\documentclass{aastex}
\usepackage{spr-astr-addons}
\usepackage{url}

\RequirePackage{color}

\newcommand{\emaila}{jameel@giki.edu.pk}
\usepackage{latexsym}
\usepackage{graphics}
\usepackage{graphicx}
\usepackage{epsfig}

\begin{document}

\title{Ground and excited states Gamow-Teller strength
distributions of iron isotopes and associated capture rates for core-collapse
simulations}

\shorttitle{GT strength distributions and capture rates on iron isotopes}

\author{Jameel-Un Nabi\altaffilmark{1}}
\affil{Faculty of Engineering Sciences, GIK Institute of
Engineering Sciences and Technology, Topi 23640, Swabi, Khyber Pakhtunkhwa,
Pakistan}

\email{\emaila}

\altaffiltext{1}{The Abdus Salam ICTP, Strada Costiera 11, 34014, Trieste, Italy}

\begin{abstract}
This paper reports on the microscopic calculation of ground and
excited states Gamow-Teller (GT) strength distributions, both in the
electron capture and electron decay direction, for $^{54,55,56}$Fe.
The associated electron and positron capture rates for these
isotopes of iron are also calculated in stellar matter. These
calculations were recently introduced and this paper is a follow-up
which discusses in detail the GT strength distributions and stellar
capture rates of key iron isotopes. The calculations are performed
within the framework of the proton-neutron quasiparticle random
phase approximation (pn-QRPA) theory. The pn-QRPA theory allows a
microscopic \textit{state-by-state} calculation of GT strength
functions and stellar capture rates which greatly increases the
reliability of the results. For the first time experimental
deformation of nuclei are taken into account. In the core of massive
stars isotopes of iron, $^{54,55,56}$Fe, are considered to be key
players in decreasing the electron-to-baryon ratio ($Y_{e}$) mainly
via electron capture on these nuclide. The structure of the
presupernova star is altered both by the changes in $Y_{e}$ and the
entropy of the core material. Results are encouraging and are
compared against measurements (where possible) and other
calculations. The calculated electron capture rates are in overall
good agreement with the shell model results. During the presupernova
evolution of massive stars, from oxygen shell burning stages till
around end of convective core silicon burning, the calculated
electron capture rates on $^{54}$Fe are around three times bigger
than the corresponding shell model rates. The calculated positron
capture rates, however, are suppressed by two to five orders of
magnitude.
\end{abstract}


\section{Introduction}
\label{intro}
Supernovae of Type Ia and Type II are considered to be the two major contributors to the
production of elements in the universe (see for example,
\cite{Arn96}). Stars in mass range $M \leq 8 M_{\odot}$ ($M_{\odot}$ denotes the
solar mass) end their life as white dwarfs. The ultimate destiny
of more massive stars is even more interesting. Once the mass of
the iron core exceeds the Chandrasekhar limit the Pauli principle
applied to electrons cannot prevent further gravitational collapse
to an even more exotic and denser form of matter than in a white
dwarf star. A violent collapse of the core is initiated which
ultimately leads to a spectacular supernova explosion (where the
luminosity of the star becomes comparable to that of an entire
galaxy containing around $10^{11}$ stars$!$). The supernova leaves
behind a compressed ball of hot neutrons -- a neutron star. As the
hot neutron star cools, any further collapse is prevented by the
Pauli principle applied to the neutrons, unless the mass is so
great that the star can become a black hole. The structure of the
progenitor star has a vital role to play in the mechanism of the
explosion. Core-collapse simulators, to-date, find it challenging
to convert the collapse into a successful explosion. A lot many
physical inputs are required at the beginning of each stage of the
entire simulation process including but not limited to collapse of
the core, formation, stalling and revival of the shock wave and
shock propagation. It is highly desirable to calculate the
presupernova stellar structure with the most reliable physical
data and inputs.

A smaller precollapse iron core mass and a lower entropy should
favor an explosion. A smaller iron core size implies less energy
loss by the shock in photodisintegrating the iron nuclei in the
overlying onion-like structure whereas a lower entropy environment
can assist to achieve higher densities for the ensuing collapse
generating a stronger bounce and in turn forming a more energetic
shock wave \cite{Bet79}. A smaller entropy also helps in reducing
the abundance of free protons thereby lowering the electron capture
rates on these free protons and resulting in a much higher value of
$Y_{e}$ at the time of bounce (at this instant of time the iron core
collapses to supranuclear density and rebounds creating a shock
wave). As a result, the shock energy
\begin{equation}
E_{S}\simeq  Y_{ef}^{10/3}(Y_{ef}-Y_{ei}),
\end{equation}
is larger for larger final lepton fraction $Y_{ef}$ (prior to
collapse) and larger difference of initial and final lepton
fraction.

Electron/positron captures and $\beta^{\pm}$-decays are amongst the
most important nuclear physics inputs that determine both the
$Y_{ef}$ and the entropy at the presupernova stage. These nuclear
weak-interaction mediated reactions directly affect the
lepton-to-baryon ratio. Further the neutrinos and antineutrinos
produced as a result of these nuclear reactions are transparent to
the stellar matter at presupernova densities and therefore assist in
cooling the core to a lower entropy state. These weak-interaction
rates are required not only in the accurate determination of the
structure of the stellar core but also bear significance in
(explosive) nucleosynthesis and element abundance calculations. As
such it is imperative to follow the evolution of the presupernova
collapse stage with a sufficiently detailed and reliable
(microscopically calculated taking into account the nuclear
structure details of the individual nuclei) nuclear reaction network
that include these weak-interaction mediated rates. Weak
interactions in presupernova stars are known to be dominated by
allowed Fermi and Gamow-Teller (GT) transitions \cite{Bet79}. The
exact r-process yields from Type II SNe are not known and are
related to the complete understanding of the mechanism of these
supernova explosions. Electron capture process is one of the
essential ingredients involved in the complex dynamics of
core-collapse supernovae and a reliable estimate of these rates can
certainly contribute in a better understanding of the explosion
mechanism.

SNe Ia are thought to be the explosions of white dwarfs that accrete
matter from binary companions and are intensively investigated for
their accurate calibration as the cosmological standard candles (one
needs to predict the SN Ia peak luminosity at least with a 10$\%$
precision). These explosions are also important for stellar
nucleosynthesis and are strongly contested to be the main producers
of Fe peak elements in the Galaxy \cite{Iwa99}. The abundance of the
Fe group, in particular of neutron rich species, is highly sensitive
to the electron captures taking place in the central layers of SNe
Ia explosions \cite{Iwa99}. These captures drive the matter to
larger neutron excesses and thus strongly influence the composition
of ejected matter and the dynamics of the explosion. It is highly
probable that electron capture occurs in the burning front driving
the matter to large neutron excess. In particular, GT
properties of nuclei in the region of medium masses around A=56 are
of special importance because they are the main constituents of the
stellar core in presupernova conditions.

The GT transition is one of the most important nuclear weak
processes of the spin-isospin ($\sigma \tau$) type. This charge
exchange transition is not only subject of interest in nuclear
physics but also of great importance in astrophysics. The GT
transitions are involved in many weak processes occurring in nuclei,
e.g. nucleosynthesis and stellar core collapse of massive stars
preceding the supernova explosion. In the early stages of the
collapse, the electron capture and $\beta$-decay are important
processes. These reactions are dominated by GT (and also by Fermi)
transitions. Electron (positron) capture rates are very sensitive to
the distribution of the GT$_{+}$ (GT$_{-}$) strength. In the
GT$_{+}$ strength, a proton is changed into a neutron (the plus sign
is for the isospin raising operator ($t_{+}$), present in the GT
matrix elements, which converts a proton into a neutron). On the
other hand the GT$_{-}$ strength is responsible for transforming a
neutron into a proton (the minus sign is for the isospin lowering
operator ($t_{-}$) which converts a neutron into a proton). The
total GT$_{+}$ strength is proportional to the electron capture
strength \cite{Roe93}.

GT$_{+}$ strength distributions on nuclei in the mass range A = 50
-- 65 have been studied experimentally mainly via (n,p) charge
exchange reactions at forward angles (e.g.
\cite{Roe93,Vet89,Elk94}). Similarly GT$_{-}$ strength distributions
were studied keenly using the (p,n) reactions (e.g.
\cite{Vet89,Rap83,And90}). It has been shown that for (p,n) and
(n,p) reactions the 0$^{0}$ cross sections for such transitions are
proportional to the squares of the matrix elements for the GT
$\beta$ decay between the same states (e.g. \cite{Goo80}). Results
of these measurements show that, in contrast to the independent
particle model, the total GT strength is quenched and fragmented
over many final states in the daughter nucleus. Both these effects
are caused by the residual interaction among the valence nucleons
and an accurate description of these correlations is essential for a
reliable evaluation of the stellar weak interaction rates due to the
strong phase space energy dependence, particularly of the stellar
electron capture rates.

Fuller, Fowler and Newman (FFN) \cite{Ful82} performed the
first-ever extensive calculation of stellar weak rates including the
capture rates, neutrino energy loss rates and decay rates for a wide
density and temperature domain. They performed this detailed
calculations for 226 nuclei in the mass range $21 \leq A \leq 60$.
They stressed on the importance of the Gamow-Teller (GT) giant
resonance strength in the capture of the electron and estimated the
GT centroids using zeroth-order ($0\hbar\omega$ ) shell model. Both
the decay and capture rates are exponentially sensitive to the
location of the GT centroid \cite{Auf96}. The location of the GT
resonance affects the stellar rates exponentially, while the total
strength affects them linearly \cite{Auf96}. Few years later
Aufderheide et al. \cite{Auf94} extended the FFN work for heavier
nuclei with A $>$ 60 and took into consideration the quenching of
the GT strength neglected by FFN. They tabulated the 90 top electron
capture nuclei averaged throughout the stellar trajectory for $0.40
\leq Y_{e} \leq 0.5$ (see Table. 25 therein). Later the experimental
results of Refs. \cite{Roe93,Vet89,Elk94,Rap83,And90} revealed the
misplacement of the GT centroid adopted in the parameterizations of
Ref. \cite{Ful82}. Since then theoretical efforts were concentrated
on the microscopic calculations of GT strength distributions and
associated weak-interaction mediated rates specially for iron-regime
nuclide. Large-scale shell model (LSSM)(e.g. \cite{Lan00}) and the
proton-neutron quasiparticle random phase approximation theory
(pn-QRPA) (e.g. \cite{Nab04}) were used extensively and with
relative success for the microscopic calculation of stellar weak
rates. Monte Carlo shell-model is an alternative to the
diagonalization method and allows calculation of nuclear properties
as thermal averages (e.g. \cite{Joh92}). However it does not allow
for detailed nuclear spectroscopy and has some restrictions in its
applications for odd-odd and odd-A nuclei.

The pn-QRPA theory is an efficient way to generate GT strength
distributions. These strength distributions constitute a primary and
nontrivial contribution to the capture rates among iron-regime
nuclide. The usual RPA was formulated for excitations in the same
nucleus. Halbleib and Sorenson \cite{Hal67} generalized this model
to describe charge-changing transitions of the type $(Z,N)
\rightarrow (Z \pm 1, N \mp 1)$ and pn-QRPA first came into
existence more than 40 years ago. The model was extended to deformed
nuclei (using Nilsson-model wave functions) by Krumlinde and
M\"{o}ller \cite{Kru84}. Extension of the model to treat odd-odd
nuclei and transitions from nuclear excited states was done by Muto
and collaborators \cite{Mut92}.

Nabi and Klapdor-Kleingrothaus used the pn-QRPA theory to calculate
the stellar weak interaction rates over a wide range of temperature
and density scale for sd- \cite{Nab99} and fp/fpg-shell nuclei
\cite{Nab04} in stellar matter (see also Ref. \cite{Nab99a}). These
included the weak interaction rates for nuclei with A = 40 to 44
(not yet calculated by shell model). Since then these calculations
were further refined with use of more efficient algorithms,
computing power, incorporation of latest data from mass compilations
and experimental values, and fine-tuning of model parameters
\cite{Nab05,Nab07,Nab07a,Nab07b,Nab08,Nab08a,Nab08b,Nab09,Nab10,Nab10a}.
There is a considerable amount of uncertainty involved in all types
of calculations of stellar weak interactions. The uncertainty
associated with the microscopic calculation of the pn-QRPA model was
discussed in detail in Ref. \cite{Nab08b}. The reliability of the
pn-QRPA calculations was discussed in length by Nabi and
Klapdor-Kleingrothaus \cite{Nab04}. There the authors compared the
measured data (half lives and B(GT$_{\pm}$) strength) of thousands
of nuclide with the pn-QRPA calculations and got fairly good
comparison.

Three key isotopes of iron, $^{54,55,56}$Fe, were selected for the
calculation of GT$_{\pm}$ strength distributions and associated
stellar electron and positron capture rates in this phase of the
project. Whereas sufficient experimental data are available for the
even-even $^{54,56}$Fe isotopes to test the model, $^{55}$Fe has
low-lying excited states which have a finite probability of
occupation in stellar conditions and a microscopic calculation of
GT$_{\pm}$  strengths from these excited states is desirable.
Aufderheide and collaborators \cite{Auf94} ranked $^{54,55,56}$Fe
amongst the most influential nuclei with respect to their importance
for the electron capture process for the early presupernova
collapse. Later Heger et al. \cite{Heg01} studied the presupernova
evolution of massive stars (of masses $15M_{\odot}, 25M_{\odot},$
and $ 40M_{\odot}$) and rated $^{54,55,56}$Fe amongst top seven
nuclei considered to be most important for decreasing $Y_{e}$ in
$15M_{\odot}$ and $40M_{\odot}$ stars. In $25M_{\odot}$ stars these
isotopes of iron were ranked as the top three key nuclei that play
the biggest role in decreasing $Y_{e}$. These isotopes of iron are
mainly responsible for decreasing the electron-to-baryon ratio
during the oxygen and silicon burning phases. Besides, $^{55}$Fe was
also found to be in the top five list of nuclei that increase
$Y_{e}$ via positron capture and electron decay during the silicon
burning phases. This paper presents the detailed analysis of the
improved microscopic calculation of GT$_{\pm}$ strength
distributions.  Details of calculation of stellar electron and
positron capture rates for these three isotopes of iron using the
pn-QRPA model are also furnished in this manuscript. Comparisons
against previous calculations are also presented. These improved
calculations were introduced recently in Ref. \cite{Nab09} where it
was reported that the betterment resulted mainly from the
incorporation of measured deformation values for these nuclei. The
idea is to present an alternate microscopic and accurate estimate of
weak interaction mediated rates to the collapse simulators which may
be used as a reliable source of nuclear physics input to the
simulation codes.

The next section briefly describes the theoretical formalism used to
calculate the GT strength distributions and the associated electron
$\&$ positron capture rates. The calculated GT$_{\pm}$ strength
distributions are presented and compared with measurements and
against other calculations in Sec. 3. The pn-QRPA calculated
electron and positron capture rates for iron isotopes
($^{54,55,56}$Fe) are presented and explored in Sec. 4. The main
conclusions of this work are finally presented in Sec. 5.

\section{Model Description}
The Hamiltonian for the pn-QRPA calculations was taken to be of the
form
\begin{equation}
H^{pn-QRPA} =H^{sp} +V^{pair} +V_{GT}^{ph} +V_{GT}^{pp},
\end{equation}
where $H^{sp}$ is the single-particle Hamiltonian (single particle
energies and wave functions were calculated in the Nilsson model,
which takes into account nuclear deformations), $V^{pair}$  is the
pairing force (pairing was treated in the BCS approximation),
$V_{GT}^{ph}$ is the particle-hole (ph) GT force, and
$V_{GT}^{pp}$  is the particle-particle (pp) GT force. The
proton-neutron residual interactions occurred as particle-hole and
particle-particle interaction. The interactions were given separable
form and were characterized by two interaction constants $\chi$  and
$\kappa$, respectively.  In this work, the values of $\chi$ and $\kappa$
was taken as 0.15 MeV and 0.07 MeV, respectively. Details of choice of these GT
strength parameters in the pn-QRPA model can be found in Refs.
\cite{Sta90,Hir93}. Other parameters
required for the calculation of weak rates are the Nilsson potential
parameters, the pairing gaps, the deformations, and the Q-values of
the reactions. Nilsson-potential parameters were taken from Ref.
\cite{Nil55} and the Nilsson oscillator constant was chosen as
$\hbar \omega=41A^{-1/3}(MeV)$ (the same for protons and neutrons).
The calculated half-lives depend only weakly on the values of the
pairing gaps \cite{Hir91}. Thus, the traditional choice of $\Delta
_{p} =\Delta _{n} =12/\sqrt{A} (MeV)$ was applied in the present
work. The deformation parameter is recently argued to be one of the
most important parameters in pn-QRPA calculations \cite{Ste04} and
as such rather than using deformations calculated from some
theoretical mass model (as used in earlier calculations of pn-QRPA
capture rates) the experimentally adopted value of the deformation
parameters for $^{54,56}$Fe, extracted by relating the measured
energy of the first $2^{+}$ excited state with the quadrupole
deformation, was taken from Raman et al. \cite{Ram87}. The incorporation of experimental
deformations lead to an overall improvement in the calculation as discussed earlier in Ref. \cite{Nab09}.
For the case of $^{55}$Fe (where measurement lacks) the deformation of the
nucleus was calculated as
\begin{equation}
\delta = \frac{125(Q_{2})}{1.44 (Z) (A)^{2/3}},
\end{equation}
where $Z$ and $A$ are the atomic and mass numbers, respectively and
$Q_{2}$ is the electric quadrupole moment taken from Ref.
\cite{Moe81}. Q-values were taken from the recent mass compilation
of Audi et al. \cite{Aud03}.

Capture rates on $^{54,55,56}Fe$ for the following two processes
mediated by charge weak interaction were calculated:
\begin{enumerate}
\item  Electron capture

\[{}_{Z}^{A} X\, +\, e^{-} \to \, {}_{Z-1}^{A} X\, +\, \nu .\]
\item  Positron capture

\[{}_{Z}^{A} X\, +\, e^{+} \, \to \, {}_{Z+1}^{A} X\, +\, \bar{\nu }.\]
\end{enumerate}
The electron capture (ec) and positron capture (pc) rates of a
transition from the $i^{th}$ state of the parent to the $j^{th}$
state of the daughter nucleus are given by
\begin{eqnarray}
\lambda ^{^{ec(pc)} } _{ij} =\left[\frac{\ln 2}{D}
\right]\left[f_{ij} (T,\rho ,E_{f} )\right] \nonumber \\
\left[B(F)_{ij}
+\left({\raise0.7ex\hbox{$ g_{A}  $}\!\mathord{\left/ {\vphantom
{g_{A}  g_{V} }} \right.
\kern-\nulldelimiterspace}\!\lower0.7ex\hbox{$ g_{V}  $}}
\right)^{2}_{eff} B(GT)_{ij} \right].
\end{eqnarray}
The value of D was taken to be 6295s \cite{Yos88}. $B_{ij}'s$ are
the sum of reduced transition probabilities of the Fermi B(F) and GT
transitions B(GT). Whereas for $^{54,56}$Fe phonon transitions
contribute, in the case of $^{55}$Fe (odd-A case) two kinds of
transitions are possible. One are the phonon transitions, where the
odd quasiparticle acts as spectator and the other is the transitions
of the odd quasiparticle itself. In the later case phonon
correlations were introduced to one-quasiparticle states in
first-order perturbation \cite{Mut89}. The $f_{ij}'s$ are the phase
space integrals. Details of the calculations of phase space
integrals and reduced transition probabilities can be found in Ref.
\cite{Nab99}. In Eq. (4) $(g_{A}/g_{V})_{eff}$ is the effective
ratio of axial and vector coupling constants and takes into account
the observed quenching of the GT strength \cite{Ost92}. For this
project $(g_{A}/g_{V})_{eff}$ was taken (from Ref. \cite{Gaa83}) as:
\begin{equation}
\left (\frac{g_{A}}{g_{V}} \right)^{2}_{eff} = 0.60 \left
(\frac{g_{A}}{g_{V}} \right)^{2}_{bare},
\end{equation}
with $(g_{A}/g_{V})_{bare}$ = -1.254 \cite{Rod06}. Interestingly,
Vetterli and collaborators \cite{Vet89} and R\"{o}nnqvist et al.
\cite{Roe93} predicted the same quenching factor of 0.6 for the RPA
calculation in the case of $^{54}$Fe when comparing their measured
strengths to RPA calculations.

The total electron (positron) capture rate per unit time per
nucleus was then calculated using
\begin{equation}
\lambda^{ec(pc)} =\sum _{ij}P_{i} \lambda _{ij}^{ec(pc)}.
\end{equation}
The summation over all initial and final states was carried out
until satisfactory convergence in the rate calculations was
achieved. Here $P_{i}$ is the probability of occupation of parent
excited states and follows the normal Boltzmann distribution. The
pn-QRPA theory allows a microscopic state-by-state calculation of
both sums present in Eq. (6). This feature of the pn-QRPA model
greatly increases the reliability of the calculated rates in stellar
matter where there exists a finite probability of occupation of
excited states.

In order to further increase the reliability of the calculated
capture rates experimental data were incorporated in the calculation
wherever possible. In addition to the incorporation of the
experimentally adopted value of the deformation parameter, the
calculated excitation energies (along with their log $ft$ values)
were replaced with an experimental one when they were within 0.5 MeV
of each other. Missing measured states were inserted and inverse and
mirror transitions (if available) were also taken into account. No theoretical
levels were replaced with the experimental ones beyond the
excitation energy for which experimental compilations had no
definite spin and/or parity. A state-by-state calculation of
GT$_{\pm}$ strength was performed for a total of 246 parent excited
states in $^{54}$Fe, 297 states for $^{55}$Fe and 266 states for
$^{56}$Fe. For each parent excited state, transitions were
calculated for 150 daughter excited states using the pn-QRPA model.
The band widths of energy states were chosen according to the
density of states to cover an excitation energy of (15-20) MeV in
parent and daughter nuclei. The summation in Eq. (6) was done to
ensure satisfactory convergence. The use of a separable interaction
assisted in the incorporation of a luxurious model space of up to 7
major oscillator shells which in turn made possible to consider
these many excited states both in parent and daughter nuclei.

\section{GT$_{\pm}$ strength distributions}
The isovector response of nuclei may be studied using the nucleon
charge-exchange reactions $(p,n)$ or $(n,p)$; by other reactions
such as $(^{3}$He,t), (d,$^{2}$He) or through heavy ion reactions.
The $0^{0}$ GT cross sections ($\Delta T =1, \Delta S =1, \Delta L
=0, \hspace{0.1cm} 0\hbar\omega$ excitations) are proportional to
the analogous beta-decay strengths. Charge-exchange reactions at
small momentum transfer can therefore be used to study beta-decay
strength distributions when beta-decay is not energetically
possible. The $(p,n)$ reactions probes the GT$_{-}$ strength
(corresponding to beta-minus decay) and the $(n,p)$ reactions gives
the strength for $\beta^{+}$-decay, i.e. GT$_{+}$ strength. The
study of $(p,n)$ reactions has the advantage over $\beta$-decay
measurements in that the GT$_{-}$  strength can be investigated over
a large region of excitation energy in the residual nucleus. On the
other hand the $(n,p)$ reactions populates only $T = T_{0}+1$ states
in all nuclei heavier than $^{3}$He. This means that other final
states (including the isobaric analog resonance) are forbidden and
GT$_{+}$ transitions can be observed relatively free of background.
The study of these reactions suggest that a reduction in the amount
of GT strength is observed relative to theoretical calculations. The
GT quenching is on the order of 30-40 $\%$ \cite{Vet89}.

In a sense both $\beta$-decay and capture rates are very sensitive
to the location of the GT$_{+}$ centroid. An $(n,p)$ experiment on a
nucleus $(Z,A)$ shows where in $(Z-1,A)$ the GT$_{+}$ centroid
corresponding to the ground state of $(Z,A)$ resides. Each excited
state of $(Z,A)$ has its own GT$_{+}$ centroid in $(Z-1,A)$ and all
of these resonances must be included in the stellar rates. We do not
have the ability to measure these resonances. Turning to theory we
see that the pioneer calculation done by FFN \cite{Ful82} had to
revert to approximations in the form of Brink's hypothesis and "back
resonances" to include all resonances in their calculation. Brink's
hypothesis states that GT strength distribution on excited states is
\textit{identical} to that from ground state, shifted \textit{only}
by the excitation energy of the state. GT back resonances are the
states reached by the strong GT transitions in the inverse process
(electron capture) built on ground and excited states. Even the
microscopic large-scale shell model calculations \cite{Lan00} had to
use the Brink assumption to include all states and resonances. On
the other hand the pn-QRPA model provides a microscopic way of
calculating the GT$_{+}$ centroid and the total GT$_{+}$ strength
for \textit{all} parent excited states and can lead to a fairly
reliable estimate of the total stellar rates.

\begin{table}[t]
\small \caption{Comparison of measured total GT$_{\pm}$ strengths
with microscopic calculations of pn-QRPA and large scale shell model
in $^{54,56}Fe$. For references see text.} \label{ta1}
 \begin{tabular}{c|cc|cc}
&  \emph{$\mathbf{^{54}Fe}$}   & &   $\mathbf{^{56}Fe}$ \\
Models & $\Sigma GT_{-} $& $\Sigma GT_{+} $ & $\Sigma GT_{-} $ & $\Sigma GT_{+} $\\
\hline pn-QRPA &  7.56 & 4.26 & 10.74 &3.71\\
Experiment  & 7.5$\pm0.7$ & 3.5$\pm0.7$ & 9.9$\pm2.4$ & 2.9$\pm0.3$\\
LSSM & 7.11 & 3.56 & 9.80 & 2.70\\\hline
\end{tabular}
\end{table}
\begin{figure}[t]
\includegraphics[width=3.3in,height=4.3in]{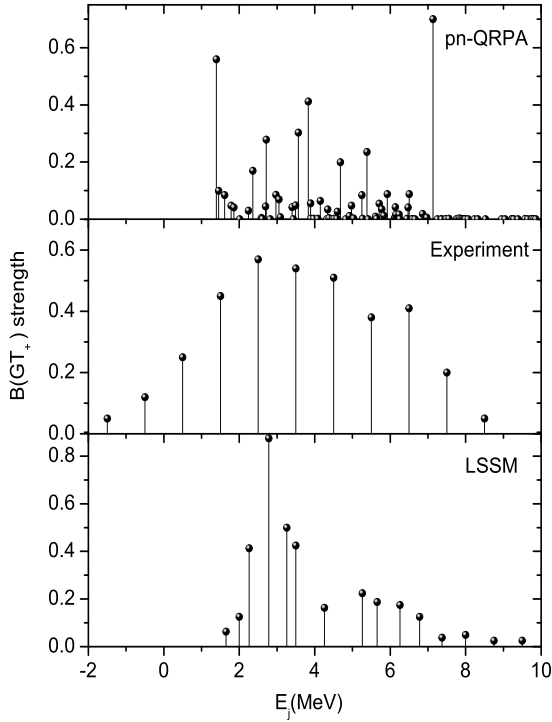}
\caption{Gamow-Teller (GT$_{+}$ ) strength distributions for
$^{54}$Fe. The upper panel shows the pn-QRPA results of GT strength
for the ground state. The middle and lower panels show the results
for the measured strength distribution \cite{Roe93} and large-scale
shell model calculations \cite{Lan00}, respectively. $E_{j}$
represents daughter states in $^{54}$Mn}
\label{fig1}
\end{figure}

As a starting point, the model was tested for the case of
$^{54,56}$Fe where measurements of total GT$_{\pm}$ strength are
available. As mentioned earlier an overall quenching factor of 0.6
\cite{Gaa83} was adopted in the calculation of GT strength in both
directions for all isotopes of iron. There is a considerable amount
of uncertainties present in all calculations of stellar rates. The
associated uncertainties in the pn-QRPA model was discussed in Ref.
\cite{Nab08b}. Keeping in mind the uncertainties present also in
measurements where various energy cutoffs are used as a reasonable
upper limit on the energy at which GT strength could be reliably
related to measured $\Delta L =0$ cross-sections, Table~\ref{ta1}
presents the comparison of the total GT$_{\pm}$ strengths against
measurements and large-scale shell model calculations \cite{Lan00}
(referred to as LSSM throughout this and proceeding sections) for
the case of $^{54,56}$Fe. Throughout this paper all energies are
given in units of MeV. For the case of $^{54}$Fe the value of the
total GT$_{+}$ was taken from Ref. \cite{Roe93} whereas the total
measured GT$_{-}$ was taken from Ref. \cite{And90}. The total
GT$_{-}$ strength calculated using the pn-QRPA model matches very
well with the measured strength (Vetterli and collaborators
\cite{Vet89} also reported a value of 7.5 $\pm$ 1.2 as the total
measured GT$_{-}$ strength for $^{54}$Fe). The calculated total
GT$_{+}$ strength lies close to the upper bound of the measured
value and are higher than the corresponding LSSM calculated
strength. For the case of $^{56}$Fe, the measured values of the
total GT$_{+}$ was taken from the latest measurement by El-Kateb and
collaborators \cite{Elk94}. The total measured GT$_{-}$ was taken
from Ref. \cite{Rap83}. In this case the calculated strengths are
relatively more enhanced as compared to LSSM calculated strengths.

As mentioned earlier the calculated rates are sensitive to the
location of the GT$_{+}$ centroid. For the case of $^{54}$Fe, the
calculated centroid of 4.06 MeV \cite{Nab09} is much higher than the
LSSM centroid of 3.78 MeV. The numbers are to be compared to the
experimental value of 3.7 $\pm $ 0.2 MeV which was calculated from
the measured data presented in Ref. \cite{Roe93}. For $^{56}$Fe, the
pn-QRPA model calculated the centroid at an excitation energy of
3.13 MeV \cite{Nab09} in the daughter nucleus $^{56}$Mn whereas LSSM
calculated it at an excitation energy of 2.60 MeV. The centroid
extracted from the experimental data of El-Kateb et al. \cite{Elk94}
comes out to be 2.9 $\pm $ 0.2 MeV. Experimental (p,n) data are also
available for $^{54,56}$Fe. According to Anderson and collaborators
\cite{And90} a large uncertainty exists in the (p,n) measurements
and the authors were able to perform measurements on $^{54}$Fe at
135 MeV with significantly better energy resolution than the earlier
measurements. Further whereas the GT strength in the discrete peaks
were extracted accurately, it was reported that the amount of GT
strength in the background and continuum remained highly uncertain.
For the sake of completeness a comparison of GT$_{-}$ centroids is
also discussed below. The centroid of the data of discrete peaks
(presented in Table I of Ref. \cite{And90}) was calculated to be
7.63 for $^{54}$Fe which is to be compared with the much lower
pn-QRPA calculated value of 5.08. For the case of $^{56}$Fe, the
calculated centroid from the data reported by Rapaport et al.
\cite{Rap83} comes out to be 8.27. The corresponding pn-QRPA
calculated value is 5.61. Despite greater experimental certainties
in the (p,n) data one notes that the pn-QRPA calculates the centroid
at much lower energies in daughter nuclei as compared to
measurements. However it is to be noted that the total strengths (in
both direction) compares very well with the measured values. Also
the location of GT$_{+}$ centroid is in very good agreement with the
measured data which is mainly responsible for the calculation of
electron capture rates on iron isotopes. For reasons not known to
the author, LSSM did not present the location of GT$_{-}$ centroids
in their paper.

\begin{figure}[t]
\includegraphics[width=3.3in,height=4.3in]{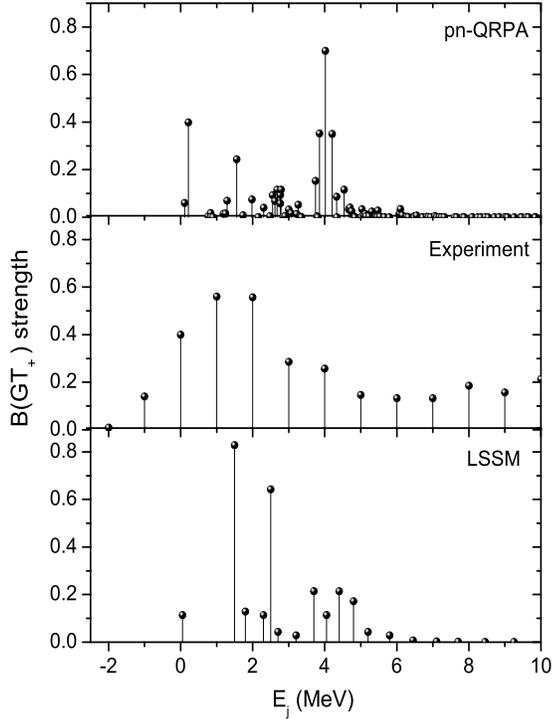}
\caption{Gamow-Teller (GT$_{+}$ ) strength distributions for
$^{56}$Fe. The upper panel shows the pn-QRPA results of GT strength
for the ground state. The middle and lower panels show the results
for the measured values \cite{Elk94} and large-scale shell model
calculations \cite{Lan00}, respectively. $E_{j}$ represents daughter
states in $^{56}$Mn.}
\label{fig2}
\end{figure}

The GT$_{+}$ strength distribution for $^{54}$Fe and $^{56}$Fe are
depicted in Fig.~\ref{fig1} and Fig.~\ref{fig2}, respectively. The
upper panel displays the calculated GT$_{+}$ distribution using the
pn-QRPA model. The middle panel shows the measured data and the
bottom panel gives the corresponding LSSM strength distributions.
One notes that an appreciable measured strength lies in the daughter
excitation energies up to around 10 MeV. The LSSM strength peaks at
much lower excitation energy. The pn-QRPA distribution spreads over
higher excitation energies akin to measured strength with a peak
around 7.1 MeV for the case of $^{54}$Fe. For the case of $^{56}$Fe
the pn-QRPA calculated spread is not as good as in case of
$^{54}$Fe. Nonetheless the pn-QRPA calculates its peak around 4 MeV
much higher in energy than the LSSM peak accumulating all the
measured higher lying strength in a narrow resonance region.

\begin{figure}[t]
\includegraphics[width=3.3in,height=4.3in]{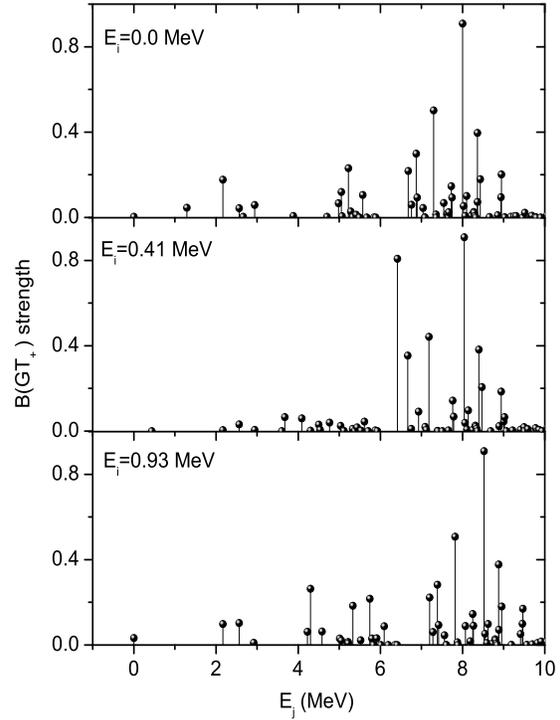}
\caption {Gamow-Teller (GT$_{+}$ ) strength distributions for
$^{55}$Fe. The upper panel shows the pn-QRPA results of GT strength
for the ground state. The middle and lower panels show the results
for the calculated first and second excited states of $^{55}$Fe.
$E_{j}$ represents daughter states in $^{55}$Mn.} \label{fig3}
\end{figure}

Fig.~\ref{fig3} shows the calculated GT$_{+}$ strength distribution
for $^{55}$Fe. Here no experimental GT distribution was available
for comparison. Instead the calculated strength for ground and first
two excited states of $^{55}$Fe (at 0.41 MeV and 0.93 MeV,
respectively) is displayed. It may be seen from Fig.3 that the GT
strength is fragmented over many daughter excited states and peaks
at relatively high excitation energies (around 8 MeV) in the
daughter $^{55}$Mn.

\begin{figure}[t]
\includegraphics[width=3.3in,height=4.3in]{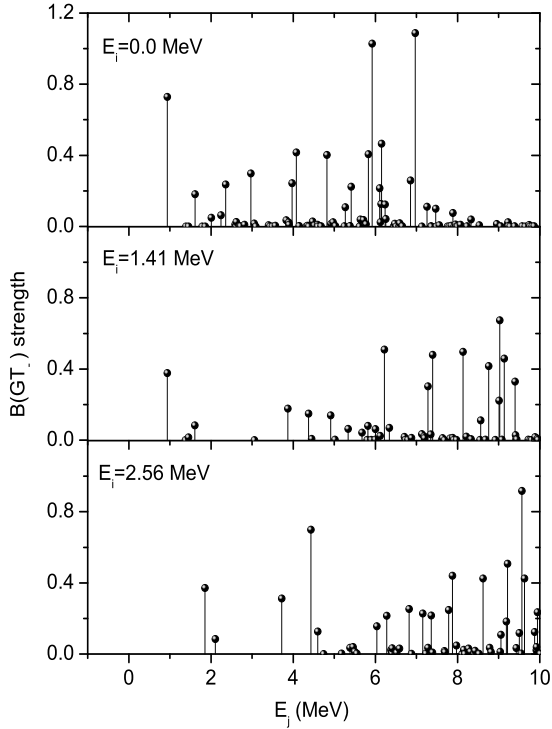}
\caption {Gamow-Teller (GT$_{-}$) strength distributions for
$^{54}$Fe. The upper panel shows the pn-QRPA results of GT strength
for the ground state. The middle and lower panels show the results
for the calculated first and second excited states of $^{54}$Fe.
$E_{j}$ represents daughter states in $^{54}$Co.} \label{fig4}
\end{figure}
\begin{figure}[t]
\includegraphics[width=3.3in,height=4.3in]{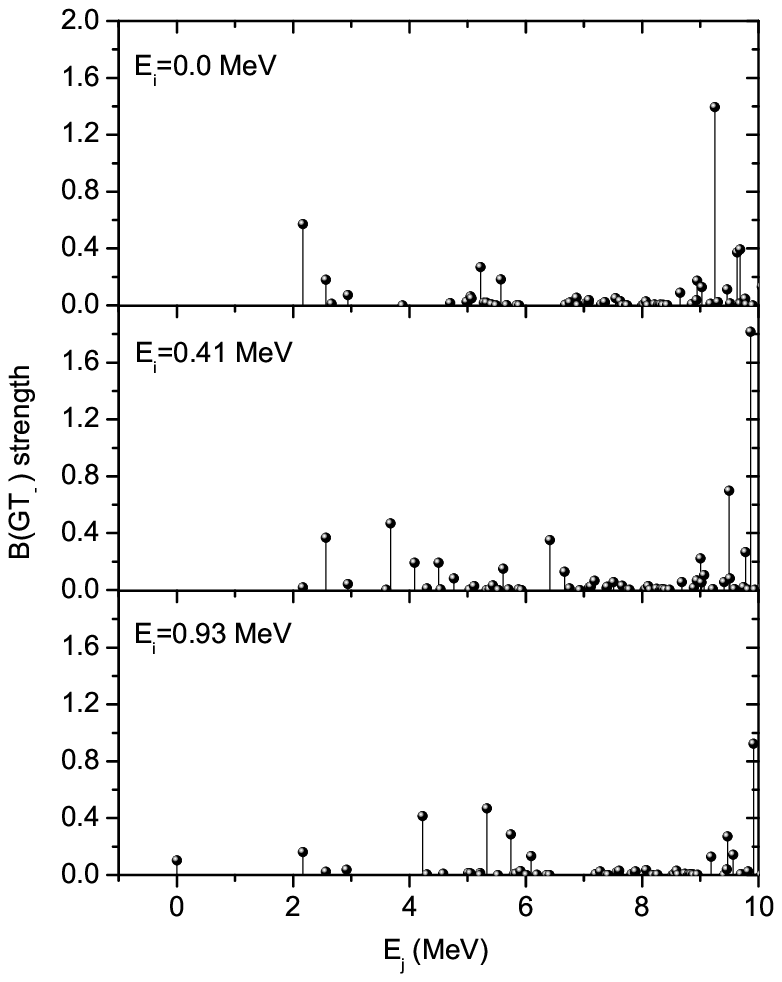}
\caption{Gamow-Teller (GT$_{-}$) strength distributions for
$^{55}$Fe. The upper panel shows the pn-QRPA results of GT strength
for the ground state. The middle and lower panels show the results
for the calculated first and second excited states of $^{55}$Fe.
$E_{j}$ represents daughter states in $^{55}$Co.} \label{fig5}
\end{figure}
\begin{figure}[t]
\includegraphics[width=3.3in,height=4.3in]{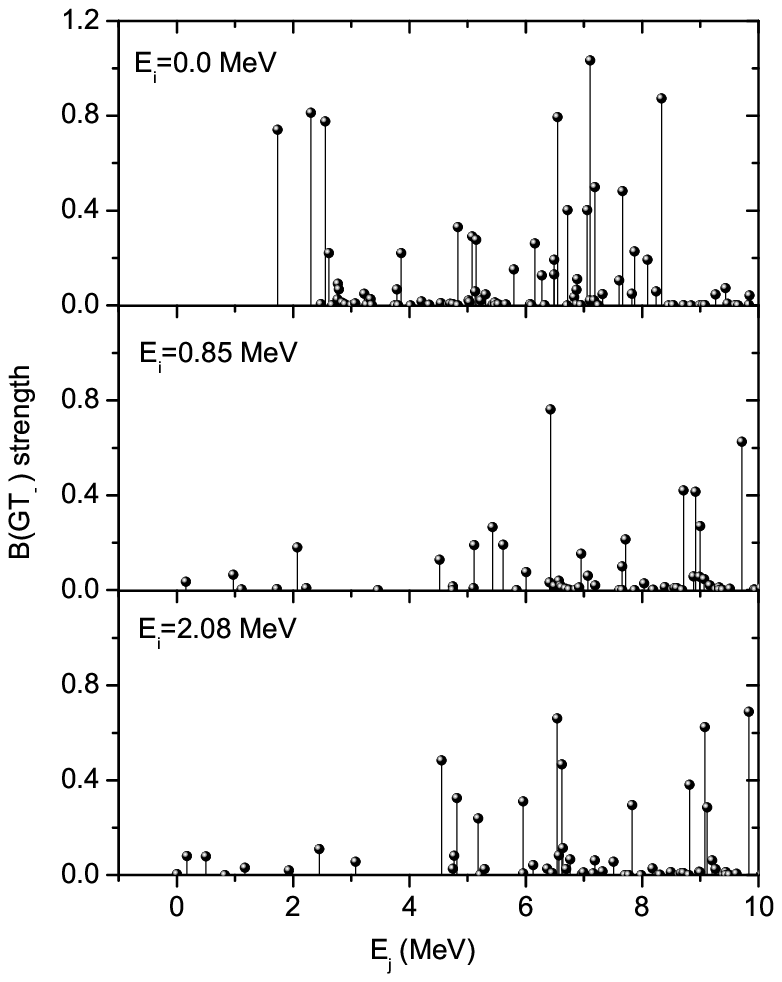}
\caption{Gamow-Teller (GT$_{-}$) strength distributions for
$^{56}$Fe. The upper panel shows the pn-QRPA results of GT strength
for the ground state. The middle and lower panels show the results
for the calculated first and second excited states of $^{56}$Fe.
$E_{j}$ represents daughter states in $^{56}$Co.} \label{fig6}
\end{figure}

Figs.~\ref{fig4},~\ref{fig5} and ~\ref{fig6} display the GT$_{-}$
strength distributions for the ground and first two calculated
excited states of $^{54}$Fe, $^{55}$Fe and $^{56}$Fe, respectively.
One notes the enhancement in the total strength in the GT$_{-}$
direction. For the case of $^{54}$Fe the peaks shift to higher
excitation energy in daughter for parent excited states.
Correspondingly the energy centroids shift to higher energy for
these excited states. For the case of $^{55}$Fe one notes that bulk
of the strength resides in a narrow resonance region around 10 MeV
in daughter. In the case of $^{56}$Fe the low-lying peaks become
considerably small in magnitude for the parent excited states.
Figs.~\ref{fig1},~\ref{fig2},~\ref{fig3},~\ref{fig4}, ~\ref{fig5},
and ~\ref{fig6} confirm that the calculated strength is fragmented
and well spread. At low temperatures and densities these low-lying
discrete strengths may very well dominate the rates and play an
important role in the state-by-state evaluation of both sums (see
Eq. (6)).

The ground and excited states strength functions calculated within
the framework of the pn-QRPA model are required in the microscopic
calculation of stellar capture rates. The ASCII files of the
GT$_{\pm}$ strength distributions for all parent excited states are
available and can be requested from the author.

\section{Electron and positron capture rates}
This section presents and explores the results of the calculated
electron and positron capture rates on $^{54,55,56}$Fe in stellar
environment and also compare the pn-QRPA capture rates with the LSSM
calculation and against the pioneering calculation of FFN
\cite{Ful82}. As discussed earlier ground and excited states GT
strengths calculated earlier contribute in the state-by-state
calculation of electron and positron capture rates (via Eq. 4).

\begin{figure}[t]
\includegraphics[width=3.3in,height=4.in]{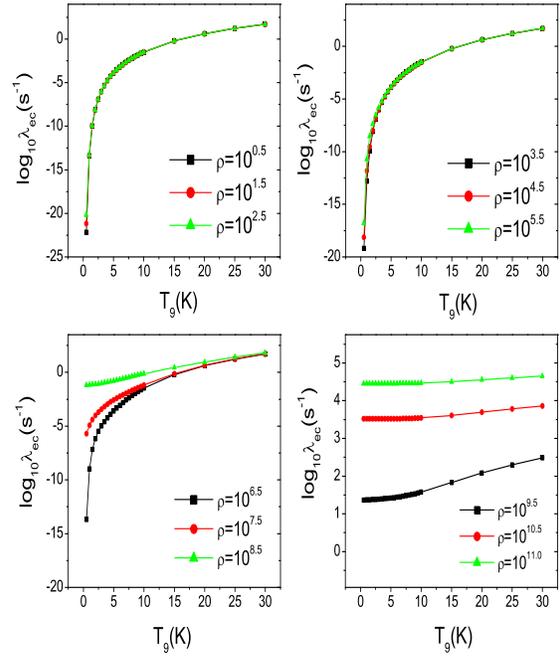}
\caption{(Color online) Electron capture rates on $^{54}$Fe, as a
function of stellar temperatures, for different selected densities.
Temperatures are given in $10^{9}$ K. Densities are given in units of $gcm^{-3}$
and log$_{10}\lambda_{ec}$ represents the log of
electron capture rates in units of $sec^{-1}$.} \label{fig7}
\end{figure}
\begin{figure}[t]
\includegraphics[width=3.3in,height=4.in]{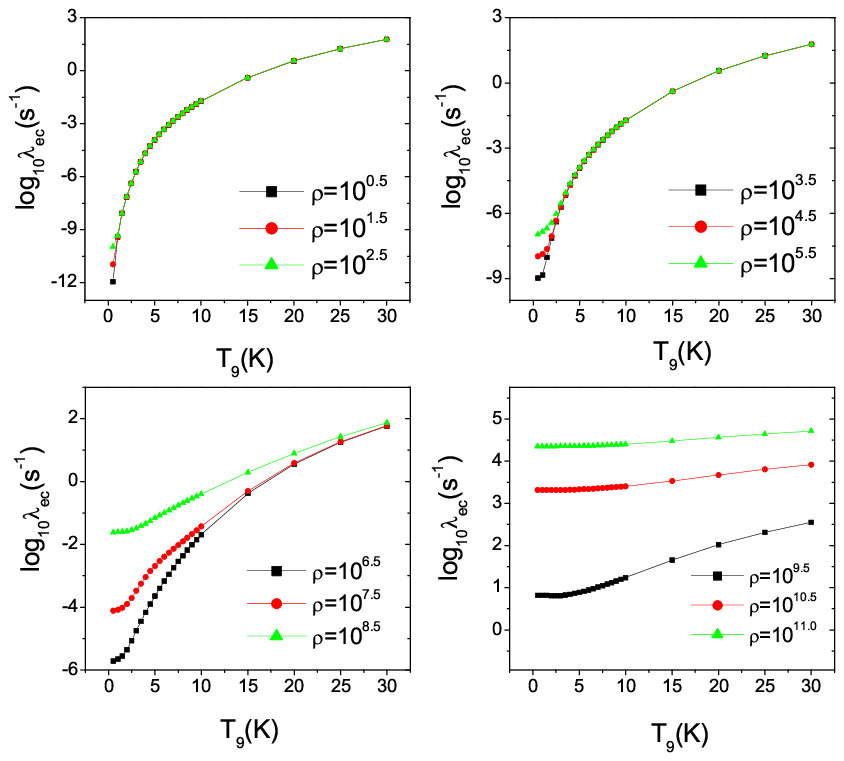}
\caption{(Color online) Same as Fig.~\ref{fig7} but for
electron capture rates on $^{55}$Fe.} \label{fig8}
\end{figure}
\begin{figure}[t]
\includegraphics[width=3.3in,height=4.in]{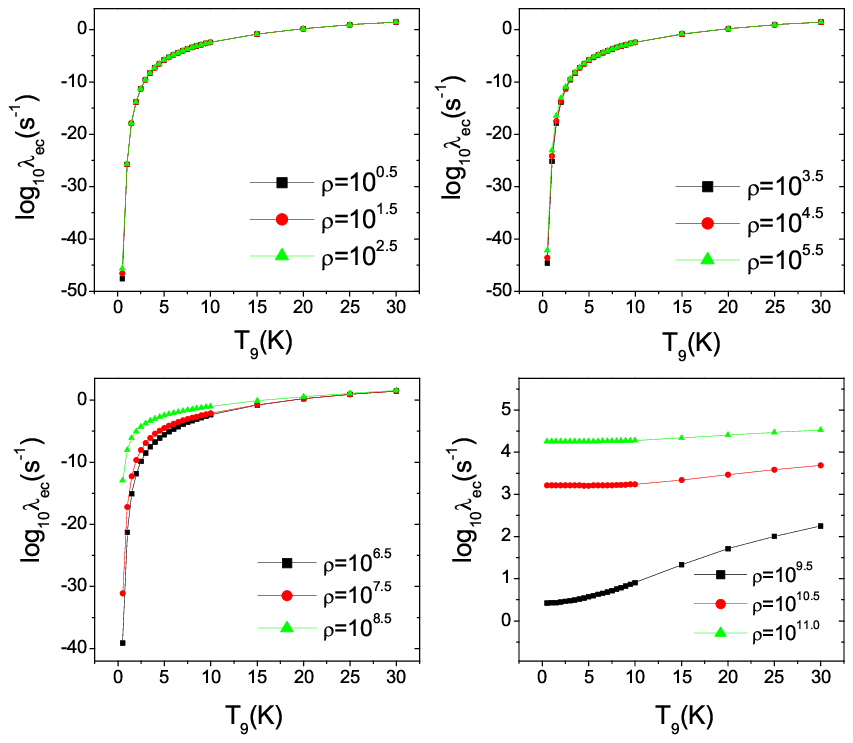}
\caption{(Color online) Same as Fig.~\ref{fig7} but for
electron capture rates on $^{56}$Fe.} \label{fig9}
\end{figure}

Figs.~\ref{fig7},~\ref{fig8} and ~\ref{fig9} show the calculated
electron capture rates on $^{54,55,56}$Fe. These figures show four
panels depicting the calculated electron capture rates at selected
temperature and density domain. The upper left panel shows the
electron capture rates in low-density region ($\rho [gcm^{-3}]
=10^{0.5}, 10^{1.5}$ and $10^{2.5}$), the upper right in medium-low
density region ($\rho [gcm^{-3}] =10^{3.5}, 10^{4.5}$ and
$10^{5.5}$), the lower left in medium-high density region ($\rho
[gcm^{-3}] =10^{6.5}, 10^{7.5}$ and $10^{8.5}$) and finally the
lower right panel depicts the calculated electron capture rates in
high density region ($\rho [gcm^{-3}] =10^{9.5}, 10^{10.5}$ and
$10^{11}$). The electron capture rates are given in logarithmic
scales (to base 10) in units of $s^{-1}$. T$_{9}$ gives the stellar
temperature in units of $10^{9}$ K. Figs.~\ref{fig7} and ~\ref{fig9}
are similar in nature depicting the electron capture rates on
even-even isotopes of iron. The rates are relatively enhanced for
the case of $^{55}$Fe by orders of magnitude (for the first three
panels). It can be seen from these figures that in the low density
regions the capture rates, as a function of temperature, are more or
less superimposed on one another. This means that there is no
appreciable change in the rates when increasing the density by an
order of magnitude. However as one goes from the medium-low density
region to high density region these rates start to 'peel off' from
one another. Orders of magnitude difference in rates are observed
(as a function of density) in high density region. For a given
density the rates increase monotonically with increasing
temperatures. For all three isotopes of iron, the calculated
electron capture rates are noticeably bigger than the competing
$\beta^{+}$ decay rates and dominate the weak rates for
charge-decreasing nuclear transitions in stellar matter.

\begin{figure}[t]
\includegraphics[width=3.3in,height=4.in]{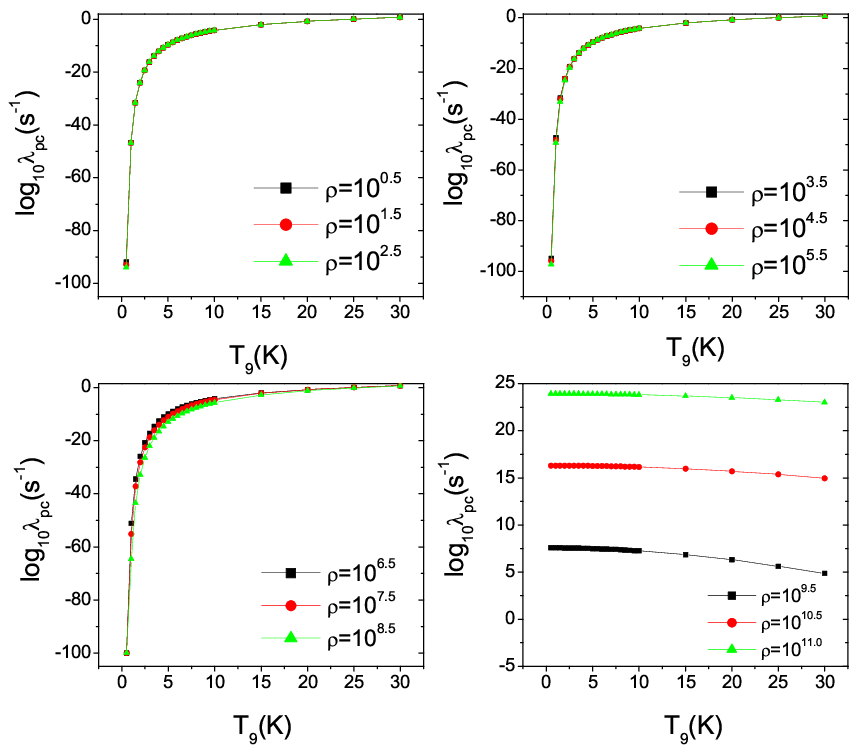}
\caption{(Color online) Same as Fig.~\ref{fig7} but for
positron capture rates on $^{54}$Fe.} \label{fig10}
\end{figure}

\begin{figure}[t]
\includegraphics[width=3.3in,height=4.in]{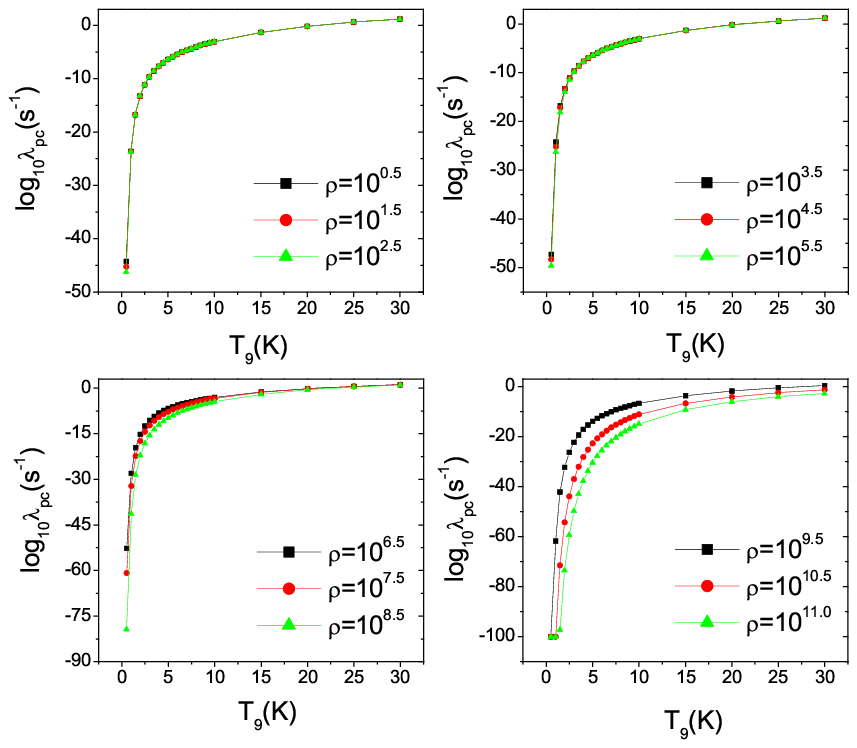}
\caption{(Color online) Same as Fig.~\ref{fig7} but for
positron capture rates on $^{55}$Fe.} \label{fig11}
\end{figure}

\begin{figure}[t]
\includegraphics[width=3.3in,height=4.in]{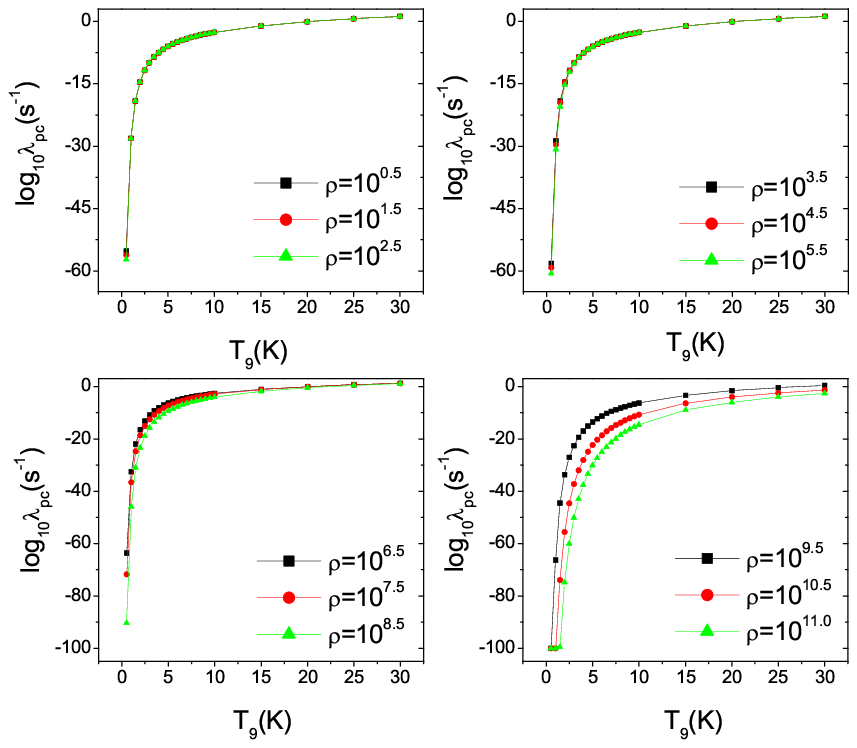}
\caption{(Color online) Same as Fig.~\ref{fig7} but for
positron capture rates on $^{56}$Fe.} \label{fig12}
\end{figure}

\begin{table}[t]
\small
\caption{The \textit{ground state} electron and positron capture rates,
$\lambda_{ec}(G)$, $\lambda_{pc}(G)$ for $^{54}Fe$ in units of
$sec^{-1}$. Given also are the ratios of the ground state capture
rates to total capture rates, $R_{ec}(G/T)$, $R_{pc}(G/T)$. The
first column gives the corresponding values of stellar density,
$\rho Y_{e}$ ($gcm^{-3}$), and temperature, $T_{9}$ (in units of
$10^{9}$ K), respectively.}
\label{ta2}
\tiny \begin{tabular}{c|cccc}
 & &\emph{$\mathbf{^{54}Fe}$} & & \\
$\mathbf{(\rho Y_{e},T_{9})}$& $\mathbf{\lambda_{ec}(G)}$ &
$\mathbf{R_{ec}(G/T)}$ &$\mathbf{\lambda_{pc}(G)}$ &
$\mathbf{R_{pc}(G/T)}$ \\\hline
(10,1)& 2.89E-18 & 1.61E-03 &    3.32E-51 & 2.83E-04 \\
(10,3)& 7.49E-08 & 4.10E-01 &    5.44E-18 & 1.71E-01 \\
(10,10)& 1.16E-02 & 4.69E-01 &    1.53E-05 & 2.69E-01 \\
(10,30)& 2.89E+00 & 5.77E-02 &    3.03E-01 & 5.22E-02 \\
\hline ($10^{3}$,1)& 4.88E-18 & 1.61E-03 &    1.97E-51 & 2.83E-04 \\
($10^{3}$,3)& 7.51E-08 & 4.10E-01 &    5.44E-18 & 1.71E-01 \\
($10^{3}$,10)& 1.16E-02 & 4.69E-01 &    1.53E-05 & 2.69E-01 \\
($10^{3}$,30)& 2.89E+00 & 5.77E-02 &    3.03E-01 & 5.22E-02 \\
\hline ($10^{7}$,1)& 3.22E-12 & 7.91E-02 &    2.97E-57 & 2.83E-04 \\
($10^{7}$,3)& 3.90E-06 & 4.96E-01 &    1.05E-19 & 1.71E-01 \\
($10^{7}$,10)& 1.45E-02 & 4.71E-01 &    1.22E-05 & 2.69E-01 \\
($10^{7}$,30)& 2.92E+00 & 5.77E-02 &    3.01E-01 & 5.22E-02 \\
\hline($10^{11}$,1)& 2.56E+04 & 1.00E+00 &    7.98E-172 & 2.83E-04 \\
($10^{11}$,3)& 2.56E+04 & 9.95E-01 &    3.49E-58 & 1.71E-01 \\
($10^{11}$,10)& 1.93E+04 & 7.12E-01 &    1.49E-17 & 2.69E-01 \\
($10^{11}$,30)& 3.14E+03 & 7.06E-02 &    4.12E-05 & 5.20E-02 \\
\hline
\end{tabular}
\end{table}
\begin{table}[t]
\small
\caption{Same as Table~\ref{ta2} but for $^{55}$Fe.}
\label{ta3}
\tiny \begin{tabular}{c|cccc}
& &\emph{$\mathbf{^{55}Fe}$} & & \\
$\mathbf{(\rho Y_{e},T_{9})}$& $\mathbf{\lambda_{ec}(G)}$ &
$\mathbf{R_{ec}(G/T)}$ &$\mathbf{\lambda_{pc}(G)}$ &
$\mathbf{R_{pc}(G/T)}$ \\\hline
(10,1) & 3.85E-10 & 9.98E-01 &    1.71E-32 & 2.48E-08 \\
(10,3) & 6.85E-07 & 3.56E-01 &    1.47E-12 & 2.01E-02 \\
(10,10) & 2.93E-03 & 1.53E-01 &    1.72E-04 & 2.35E-01 \\
(10,30) & 1.51E+00 & 2.56E-02 &    3.61E-01 & 2.34E-02 \\
\hline ($10^{3}$,1) & 6.51E-10 & 9.98E-01 &    1.01E-32 & 2.48E-08\\
($10^{3}$,3) & 6.86E-07 & 3.56E-01 &    1.47E-12 & 2.01E-02\\
($10^{3}$,10)& 2.93E-03 & 1.53E-01 &    1.72E-04 & 2.35E-01\\
($10^{3}$,30) & 1.51E+00 & 2.56E-02 &    3.62E-01 & 2.34E-02 \\
\hline ($10^{7}$,1) & 1.22E-05 & 9.99E-01 &    1.53E-38 & 2.48E-08 \\
($10^{7}$,3) & 2.90E-05 & 4.28E-01 &    2.83E-14 & 2.01E-02 \\
($10^{7}$,10)& 3.68E-03 & 1.54E-01 &    1.37E-04 & 2.35E-01 \\
($10^{7}$,30)& 1.52E+00 &2.56E-02 &    3.59E-01 & 2.34E-02\\
\hline($10^{11}$,1)& 2.25E+04 & 9.92E-01 &    4.11E-153 & 2.48E-08 \\
($10^{11}$,3) & 1.83E+04 & 8.07E-01 &    9.44E-53 & 2.01E-02\\
($10^{11}$,10)& 8.43E+03 & 3.35E-01 &    1.67E-16 & 2.35E-01\\
($10^{11}$,30)& 2.08E+03 & 4.03E-02 &    4.94E-05 & 2.33E-02 \\
\hline
\end{tabular}
\end{table}

\begin{table}[t]
\small
\caption{Same as Table~\ref{ta2} but for $^{56}$Fe.}
\label{ta4}
\tiny \begin{tabular}{c|cccc}
 & &\emph{$\mathbf{^{56}Fe}$} & & \\
$\mathbf{(\rho Y_{e},T_{9})}$& $\mathbf{\lambda_{ec}(G)}$ &
$\mathbf{R_{ec}(G/T)}$ &$\mathbf{\lambda_{pc}(G)}$ &
$\mathbf{R_{pc}(G/T)}$ \\\hline
(10,1)& 8.33E-27 & 4.70E-01 & 6.49E-37 & 7.01E-08\\
(10,3)& 1.59E-10 & 5.69E-01 & 2.26E-13 & 1.77E-02\\
(10,10)& 1.60E-03 & 4.07E-01 & 3.80E-04 & 3.21E-01\\
(10,30)& 1.52E+00 & 5.33E-02 & 9.90E-01 & 5.77E-02\\
\hline ($10^{3}$,1)& 1.41E-26 & 4.70E-01 & 3.84E-37 & 7.01E-08\\
($10^{3}$,3)& 1.59E-10 & 5.69E-01 & 2.25E-13 & 1.77E-02\\
($10^{3}$,10) & 1.60E-03 & 4.07E-01 & 3.80E-04 & 3.21E-01\\
($10^{3}$,30)& 1.52E+00 & 5.33E-02 & 9.91E-01 & 5.77E-02\\
\hline ($10^{7}$,1)& 9.30E-21 & 4.77E-01 & 5.82E-43 & 7.01E-08\\
($10^{7}$,3)& 8.28E-09 & 5.71E-01 & 4.34E-15 & 1.77E-02\\
($10^{7}$,10)& 2.01E-03 & 4.07E-01 & 3.03E-04 & 3.20E-01\\
($10^{7}$,30)& 1.53E+00 & 5.33E-02 & 9.83E-01 & 5.77E-02\\
\hline($10^{11}$,1)& 1.80E+04 & 1.00E+00 & 1.56E-157 & 7.01E-08\\
($10^{11}$,3) & 1.74E+04 & 9.64E-01 & 1.45E-53 & 1.77E-02\\
($10^{11}$,10)& 1.14E+04 & 6.03E-01 & 3.70E-16 & 3.18E-01\\
($10^{11}$,30)& 2.13E+03 & 6.35E-02 & 1.35E-04 & 5.74E-02\\
\hline
\end{tabular}
\end{table}

The effects of positron capture are estimated to be smaller for
stars in the mass range $(10 \leq M_{\odot} \leq 40)$ and could be
greater for more massive stars \cite{Heg01}. The positron capture
rates are normally bigger than the competing $\beta^{-}$ rates for
all isotopes of iron specially at high temperatures.
Figs.~\ref{fig10},~\ref{fig11} and ~\ref{fig12} again show four
panels depicting the calculated positron capture rates at selected
temperature and density domain. The upper left panel shows the
positron capture rates in low-density region ($\rho [gcm^{-3}]
=10^{0.5}, 10^{1.5}$ and $10^{2.5}$), the upper right in medium-low
density region ($\rho [gcm^{-3}] =10^{3.5}, 10^{4.5}$ and
$10^{5.5}$), the lower left in medium-high density region ($\rho
[gcm^{-3}] =10^{6.5}, 10^{7.5}$ and $10^{8.5}$) and finally the
lower right panel depicts the calculated positron capture rates in
high density region ($\rho [gcm^{-3}] =10^{9.5}, 10^{10.5}$ and
$10^{11}$). The positron capture rates are given in logarithmic
scales in units of $s^{-1}$. T$_{9}$ gives the stellar temperature
in units of $10^{9}$ K. One should note the order of magnitude
enhancement in positron capture rates as the stellar temperature
increases. Around presupernova temperatures the positron capture
rates are very slow as compared to the corresponding electron
capture rates. When the temperature of the stellar core increases
further the positron capture rates shoot up. The positron capture
rates decrease with increasing densities, in contrast to the
electron capture rates which increase as density increases. As
temperature rises or density lowers (the degeneracy parameter is
negative for positrons), more and more high-energy positrons are
created leading in turn to higher capture rates. It is worth
mentioning that the positron capture rates are very small numbers
and can change by orders of magnitude by a mere change of 0.5 MeV,
or less, in parent or daughter excitation energies and are more
reflective of the uncertainties in the calculation of energies.

In order to present an estimate of the contribution of excited
states to the total capture rate, the ground state capture rate and
the ratio of this rate to the total rate were computed at selected
points of stellar temperature and density. These contributions vary
for different isotopes and lead to interesting consequences as presented
below.

Tables~\ref{ta2},~\ref{ta3}, and ~\ref{ta4} show the contribution of excited states in the
calculation of total capture rates on $^{54,55,56}$Fe. In all tables
the capture rates and the ratios are calculated at selected points
of density and temperature shown in first column. Within the
parenthesis the first number gives the density in units of
$gcm^{-3}$ and the second number denotes the stellar temperature in
units of $10^{9}$ K. In the second column $\lambda_{ec}(G)$ gives
the ground state electron capture rates whereas $R_{ec}(G/T)$
denotes the ratios of the ground-state electron capture rate to the
total rate. The ground state positron capture rates,
$\lambda_{pc}(G)$, and the ratios of the ground-state positron
capture rate to the total rate, $R_{pc}(G/T)$, are given in the
fourth and fifth column, respectively. All capture rates are given
in units of $sec^{-1}$.

The results for $^{54}$Fe are very interesting and are depicted in
Table~\ref{ta2}. The total electron capture rate increases with
increasing temperatures and densities. At low temperatures and
densities (T$_{9} =1$ and $\rho Y_{e} \leq 10^{3} gcm^{-3}$) one
notes that almost all of the contribution to the total electron
capture rate comes from the excited states. At a first glance this
might look odd. To explain this result one has to go back to Eq. (4)
which tells that the partial capture rate is a product of three
factors: a constant, phase space integrals (which are functions of
temperature and density) and reduced transition probabilities. For a
particular parent state ($i$) all such transitions are summed over
daughter states ($j$) and then multiplied by the probability of
occupation of that parent excited state. These partial capture rates
are finally summed over all parent excited states to calculate the
total capture rate as a function of stellar temperature and density
(Eq. (6)). For the given physical conditions, the ground state
electron capture rate $ \sim 10^{-18} s^{-1}$. Probability of
occupation of ground state is essentially $1$ at such low
temperatures. However for the first excited state (at 1.41 MeV) the
partial rate increases roughly by 10 orders of magnitude to around
$10^{-8} s^{-1}$. This increase is attributed to a simultaneous
increase in phase space (= $Q+E_{i}-E_{j}$ where $Q$ = -0.697 MeV
\cite{Aud03}) as well as an increase in the total GT$_{+}$ strength
from 4.26 units to 5.12 units (refer to Table 2 of Ref.
\cite{Nab09}). Even after multiplying by the much smaller
probability of occupation of parent excited state ($ \sim 10^{-8}$)
the partial capture rate from the first excited state is around 3
orders of magnitude bigger than that from ground state at low
temperatures. As density increases the phase space from excited
states decreases resulting in a larger contribution to total rate
from ground state. In the limiting case of densities around $10^{11}
gcm^{-3}$ all the contribution to the total rate comes from ground
state. At higher temperatures ($3 \leq T_{9} \leq 10$) and densities
($ \leq 10^{7} gcm^{-3}$) the contribution of ground state to total
rate is roughly 50$\%$. At still higher temperatures (T$_{9} = 30$)
the ground state contributes around 6--7$\%$ at all densities. At
T$_{9} = 30$, the partial capture rates are significant also for
higher lying excited states (as probability of occupation of these
high-lying states then becomes appreciably high) and contribute
effectively to the total capture rate. For the corresponding
contribution of excited states to total positron capture rate, one
sees that these contributions are essentially dependent on stellar
temperatures and independent of stellar densities. Further one notes
that the contribution of excited states dominate in total positron
capture rates. These features are reflective of the fact that the
positron capture rates are dominated by phase space integrals. For a
given temperature the total positron capture rate increases by a
constant factor of the corresponding ground state rate independent
of the densities. At low temperatures the ground state contribution
is almost negligible for reasons mentioned above. As the stellar
temperature increases so does the ground state contribution until it
reaches a maximum around T$_{9} =10$ of the order of 27$\%$. At
T$_{9} =30$ roughly 95$\%$ of the contribution comes from excited
states.

For the case of $^{55}$Fe the story is much different
(Table~\ref{ta3}). Even though the first excited state is at 0.41
MeV (as against 1.41 MeV for the case of $^{54}$Fe) resulting in a
much bigger probability of occupation, the partial electron capture
rate from first excited state is around 7 orders of magnitude
smaller than the corresponding ground state electron capture rate $
\sim 10^{-10} s^{-1}$ (the total GT$_{+}$ strength decreases for the
first excited state from 4.68 units to 4.43 units \cite{Nab09}). As
a result the total electron capture rate is determined almost
entirely from ground state rate at low temperatures. This is in
sharp contrast to the previous case. The low-lying excited states do
play their role at higher temperatures and densities. The behavior
of ground state contribution to the total electron capture rate is
as per expectation and decreases appreciably as the stellar
temperature increases. Again at T$_{9} =30$ almost 97$\%$ of the
contribution is from the parent excited states at all densities akin
to the case of $^{54}$Fe. The excited states contribution to the
total positron capture rate follows a similar trend as in previous
case with a maximum contribution of around 25$\%$ from ground state
at T$_{9} =10$. At T$_{9} =3$ the ground state contributes $ \sim
2\% $ (as against 17$\%$ in the case of $^{54}$Fe). At much higher
temperature (T$_{9} \sim 30$) this contribution is again around
2$\%$.

The excited states contribution to total electron capture rate on
$^{56}$Fe at low temperatures and densities falls roughly in between
the two extreme cases of $^{54}$Fe and $^{55}$Fe (Table~\ref{ta4}).
Here roughly 50$\%$ contribution comes from the excited states at
low temperatures and densities. Example giving at T$_{9} =1$ and
$\rho Y_{e} \leq 10 gcm^{-3}$, the total ground state electron
capture rate is around $10^{-27} s^{-1}$. The contribution from
first excited state (0.85 MeV) is around $10^{-22} s^{-1}$. This is
again attributed partly to the increased total GT$_{+}$ strength of
5.15 units as against 3.71 units (for ground state) \cite{Nab09}.
This rate when multiplied with the occupation probability of around
$10^{-5}$ gives a partial electron capture rate of roughly equal
magnitude as that from ground state. The excited states contribution
increases at higher temperatures with roughly 95$\%$ of the
contribution coming from parent excited states at T$_{9} =30$. One
also notes that the ground state contribution to the total electron
capture rates increases at $\rho Y_{e} \sim 10^{11} gcm^{-3}$.
Specially at low temperatures it is almost totally dictated by
ground state rates. This is because ground and excited state rates
are of comparable magnitudes at this density. However the
probability of occupation of excited states is smaller by orders of
magnitude at lower temperatures. The excited states contribution to
total positron capture rate follows a similar trend as in previous
cases with a maximum contribution of around 30$\%$ from ground state
at T$_{9} =10$. Tables~\ref{ta2},~\ref{ta3}, and ~\ref{ta4}
highlight the contribution of partial capture rates from parent
excited states to the total capture rate and show disparate behavior
of these contributions for all three cases. This analysis further
stresses on the fact that a microscopic calculation of GT strength
function from excited states is indeed required for a reliable
estimate of the total capture rates.

One relevant question would then be to ask how the calculated rates
compare with other prominent calculations. For the comparison two
important calculations of capture rates were taken into
consideration: the pioneer calculation of FFN \cite{Ful82} which is
still used in many simulation codes and the microscopic calculation
of LSSM.

\begin{table*}
\scriptsize
\caption{Ratios of calculations of
electron capture rates on $^{54}Fe$ at different selected densities
and temperatures. QRPA implies the reported rates whereas SM and FFN
denote rates calculated by Ref. \cite{Lan00} and Ref. \cite{Ful82},
respectively.\label{ta5}}
\begin{tabular}{ccccccccc}   $T_{9}$ & QRPA/SM & QRPA/FFN
&QRPA/SM & QRPA/FFN & QRPA/SM & QRPA/FFN & QRPA/SM & QRPA/FFN
\\
        & $10gcm^{-3}$ & $10gcm^{-3}$ & $10^{3}gcm^{-3}$ & $10^{3}gcm^{-3}$ &
        $10^{7}gcm^{-3}$ & $10^{7}gcm^{-3}$ & $10^{11}gcm^{-3}$& $10^{11}gcm^{-3}$\\\hline
1&     4.49E+00& 1.04E-01&    4.49E+00& 1.05E-01&    3.05E+00&  1.31E-01&    8.59E-01&    3.00E-01\\
3&     2.50E+00& 1.93E-01&    2.50E+00& 1.93E-01&    2.38E+00&  2.36E-01&    8.59E-01&    2.99E-01\\
10&    9.95E-01& 4.36E-01&    9.93E-01& 4.36E-01&    9.95E-01&  4.37E-01&    8.45E-01&    2.99E-01\\
30&    1.14E+00& 4.00E-01&    1.14E+00& 4.00E-01&    1.14E+00&
4.00E-01&    1.05E+00&    3.62E-01\\ \hline
\end{tabular}
\end{table*}

The comparisons are presented in a tabular form. Tables~\ref{ta5},~\ref{ta6},
and ~\ref{ta7} show the comparison of calculated electron capture rates with
those of FFN and LSSM for $^{54}$Fe, $^{55}$Fe and $^{56}$Fe,
respectively. Here the ratios of the calculated electron capture
rates to those of FFN and LSSM are presented at selected temperature
and density points. For the case of $^{54}$Fe, the calculated
electron capture rates are in good agreement with the LSSM results
specially at high temperatures and densities.

At low temperatures and densities the reported electron capture
rates are bigger than the LSSM capture rates by around a factor of
four (Table~\ref{ta5}). During the oxygen shell burning and silicon core
burning of massive stars (as per the simulation results of
\cite{Heg01}) the pn-QRPA calculated rates are enhanced roughly by a
factor of three as compared to LSSM rates. Whereas the individual
discrete transitions between initial and final states matter at low
temperatures and densities, it is the total GT strength that counts
at high temperatures and densities. It is again reminded that
Brink's hypothesis is not assumed in this calculation (which was
adopted in LSSM calculation of weak rates). (Also see
the discussion on contribution of excited state partial rates
above.) The ground state pn-QRPA centroid is placed at much higher
energy in $^{54}$Mn as compared to LSSM centroid \cite{Nab09}.
However, pn-QRPA calculates a much bigger total strength. FFN rates
are up to around an order of magnitude enhanced compared to pn-QRPA
rates at low temperatures and densities. FFN did not take into
effect the process of particle emission from excited states and
their parent excitation energies extended well beyond the particle
decay channel. These high lying excited states begin to show their
effect specially at high temperatures and densities. Also one notes
that FFN neglected the quenching of the total GT strength. The LSSM
electron capture rates were smaller than the FFN rates by, on
average, an order of magnitude \cite{Heg01}.

\begin{table*}
\scriptsize
\caption{Same as Table 5, but for electron capture
on $^{55}Fe$.\label{ta6}}
\begin{tabular}{ccccccccc}   $T_{9}$ & QRPA/SM & QRPA/FFN
&QRPA/SM & QRPA/FFN & QRPA/SM & QRPA/FFN & QRPA/SM & QRPA/FFN
\\
        & $10gcm^{-3}$ & $10gcm^{-3}$ & $10^{3}gcm^{-3}$ & $10^{3}gcm^{-3}$ &
        $10^{7}gcm^{-3}$ & $10^{7}gcm^{-3}$ & $10^{11}gcm^{-3}$& $10^{11}gcm^{-3}$\\\hline
1 &  1.01E+00&    1.00E+00&   1.01E+00&    9.98E-01&    1.02E+00&   1.01E+00&    9.04E-01&   2.04E-01\\
3 &  1.45E+00&    3.72E-01&   1.44E+00&    3.72E-01&   1.44E+00&    3.58E-01 &   8.91E-01&   2.04E-01\\
10 & 1.12E+00&    1.12E-01&   1.12E+00&    1.12E-01&   1.12E+00&    1.12E-01&   9.23E-01&   2.19E-01\\
30&  1.72E+00&    3.50E-01&   1.72E+00&    3.51E-01&  1.72E+00 &
3.50E-01 &  1.44E+00 &   3.53E-01\\ \hline
\end{tabular}
\end{table*}

The electron capture rates on $^{55}$Fe are most effective during
the oxygen shell burning till around the ignition of the first stage
of convective silicon shell burning of massive stars \cite{Heg01}.
For the corresponding temperatures and densities, the pn-QRPA rates
are in very good agreement with the LSSM rates at all temperatures
and densities (Table~\ref{ta6}). This excellent agreement is attributed to
the fact that the total rates for $^{55}$Fe are commanded by ground
state contribution (see Table~\ref{ta3}) and both pn-QRPA and LSSM perform
a microscopic calculation of the ground-state GT strength function. The comparison
with FFN is good at low temperatures only (T$_{9} \sim 1$). At
higher temperatures FFN rates are bigger than the LSSM and pn-QRPA
rates by an order of magnitude for reasons mentioned before.

\begin{table*}
\scriptsize
\caption{Same as Table 5, but for electron capture
on $^{56}Fe$.\label{ta7}}
\begin{tabular}{ccccccccc}   $T_{9}$ & QRPA/SM & QRPA/FFN
&QRPA/SM & QRPA/FFN & QRPA/SM & QRPA/FFN & QRPA/SM & QRPA/FFN
\\
        & $10gcm^{-3}$ & $10gcm^{-3}$ & $10^{3}gcm^{-3}$ & $10^{3}gcm^{-3}$ &
        $10^{7}gcm^{-3}$ & $10^{7}gcm^{-3}$ & $10^{11}gcm^{-3}$& $10^{11}gcm^{-3}$\\\hline
1 &  8.87E-01&    2.77E-01&   8.87E-01&    2.78E-01&   1.01E+00&    3.37E-01&  1.13E+00&    3.37E-01\\
3  & 1.12E+00&    2.63E-01&   1.12E+00&    2.63E-01&   1.17E+00&    2.77E-01 & 1.10E+00 &   3.33E-01\\
10 & 1.02E+00&    4.80E-01 &  1.02E+00&    4.80E-01 &  1.02E+00&    4.80E-01 & 1.07E+00&    3.25E-01\\
30 & 1.29E+00&    5.04E-01&   1.29E+00&    5.04E-01&   1.29E+00&
5.02E-01&  1.37E+00&    4.14E-01\\ \hline
\end{tabular}
\end{table*}

The average comparison of calculated electron capture rates is again
very good against the LSSM rates for the case of $^{56}$Fe
(Table~\ref{ta7}). However at low temperatures and densities the LSSM
electron capture rates is around 10$\%$ enhanced as compared to
pn-QRPA numbers. As mentioned above the comparison of the reported
and LSSM electron capture rates may be traced back to the
calculation of excited state partial capture rates (Tables~\ref{ta2},~\ref{ta3}, and ~\ref{ta4}).
One notes that at T$_{9} =1$ and $\rho Y_{e} \leq 10 gcm^{-3}$ the
electron capture rates were dominated by excited state contributions
for the case of $^{54}$Fe. For the case of $^{55}$Fe it was entirely
dominated by ground state contribution and the case of $^{56}$Fe had
a 50-50 contribution. Correspondingly the comparison with LSSM rates
is excellent for the case of $^{55}$Fe. The pn-QRPA rates are
enhanced by a factor of 4 for $^{54}$Fe and some fluctuations  in
rate comparison can be seen for the case of $^{56}$Fe. The electron
capture rates on $^{56}$Fe are very important for the pre-supernova
phase of massive stars. FFN rates are again enhanced by up to an
order of magnitude as compared to pn-QRPA and LSSM rates.

\begin{table*}
\scriptsize
\caption{Same as Table 5, but for positron
capture on $^{54}Fe$.\label{ta8}}
\begin{tabular}{ccccccccc}   $T_{9}$ & QRPA/SM & QRPA/FFN
&QRPA/SM & QRPA/FFN & QRPA/SM & QRPA/FFN & QRPA/SM & QRPA/FFN
\\
        & $10gcm^{-3}$ & $10gcm^{-3}$ & $10^{3}gcm^{-3}$ & $10^{3}gcm^{-3}$ &
        $10^{7}gcm^{-3}$ & $10^{7}gcm^{-3}$ & $10^{11}gcm^{-3}$& $10^{11}gcm^{-3}$\\\hline
1 &  3.50E-05&    3.53E-05&   3.50E-05&    3.53E-05&   3.50E-05&    3.52E-05&  9.98E-01&    9.98E-01\\
3  & 2.91E-02&    3.18E-02&   2.91E-02&    3.17E-02&   2.91E-02&    3.18E-02 & 2.90E-02 &   3.18E-02\\
10 & 6.97E-01&    8.47E-01 &  6.97E-01&    8.47E-01 &  6.97E-01&    8.43E-01 & 6.89E-01&    8.24E-01\\
30 & 7.59E+00&    8.51E-01&   7.60E+00&    8.53E-01&   7.62E+00&    8.55E-01&  7.41E+00&    8.38E-01\\
\hline
\end{tabular}
\end{table*}

Tables~\ref{ta8},~\ref{ta9}, and ~\ref{ta10} show the corresponding comparison for the
calculated positron capture rates. Again the pn-QRPA calculated
positron capture rates are compared with the LSSM and FFN rates. It
is reminded that the positron capture rates are smaller than the
corresponding electron capture rates by orders of magnitude and
these small numbers can change appreciably by a mere change of 0.5
MeV in phase space and are actually more reflective of the
uncertainties in calculation of the energy eigenvalues (for both
parent and daughter states). Further it is also evident from Tables~\ref{ta8},~\ref{ta9},
and ~\ref{ta10} that positron capture rates have a dominant
contribution from parent excited states (the ground state
contributes at the maximum by one third). Looking at Table~\ref{ta8} for
the case of $^{54}$Fe one notes that the pn-QRPA calculated positron
capture rates are suppressed by up to 5 orders of magnitude as
compared to LSSM and FFN rates at low temperatures and densities.
The comparison improves as the temperature and density increases. At
T$_{9} =30$ the reported rates are in fact enhanced by a factor of 8
as compared to LSSM calculated rates and are in reasonable agreement
with FFN rates.

\begin{table*}
\scriptsize
\caption{Same as Table 5, but for positron
capture on $^{55}Fe$.\label{ta9}}
\begin{tabular}{ccccccccc}   $T_{9}$ & QRPA/SM & QRPA/FFN
&QRPA/SM & QRPA/FFN & QRPA/SM & QRPA/FFN & QRPA/SM & QRPA/FFN
\\
        & $10gcm^{-3}$ & $10gcm^{-3}$ & $10^{3}gcm^{-3}$ & $10^{3}gcm^{-3}$ &
        $10^{7}gcm^{-3}$ & $10^{7}gcm^{-3}$ & $10^{11}gcm^{-3}$& $10^{11}gcm^{-3}$\\\hline
1 &  1.85E-03&    1.57E-03&  1.85E-03&    1.57E-03&  1.85E-03&    1.57E-03&  9.98E-01&    9.98E-01\\
3  & 4.73E-01&    4.12E-02&  4.73E-01&    4.12E-02&  4.72E-01&    4.06E-02&  4.69E-01&    4.06E-02\\
10 & 1.45E+00&    1.34E-01 & 1.45E+00&    1.34E-01&  1.44E+00&    1.34E-01&  1.40E+00&    1.31E-01\\
30 & 1.15E+01&    8.39E-01&  1.15E+01&    8.41E-01&  1.15E+01&    8.39E-01&  1.12E+01&    8.24E-01\\
\hline
\end{tabular}
\end{table*}

The comparison follows a similar trend for the case of $^{55}$Fe
(Table~\ref{ta9}). Here the pn-QRPA calculated rates are suppressed
by as much as 3 orders of magnitude compared with the other
calculations at low temperatures.  The comparison improves as the
density of stellar core stiffens. The reported rates are enhanced at
higher temperatures by as much as an order of magnitude as compared
to LSSM numbers. FFN rates are again in reasonable agreement with
reported rates at T$_{9} =30$.

\begin{table*}
\scriptsize
\caption{Same as Table 5, but for positron
capture on $^{56}Fe$.\label{ta10}}
\begin{tabular}{ccccccccc}   $T_{9}$ & QRPA/SM & QRPA/FFN
&QRPA/SM & QRPA/FFN & QRPA/SM & QRPA/FFN & QRPA/SM & QRPA/FFN
\\
       & $10gcm^{-3}$ & $10gcm^{-3}$ & $10^{3}gcm^{-3}$ & $10^{3}gcm^{-3}$ &
        $10^{7}gcm^{-3}$ & $10^{7}gcm^{-3}$ & $10^{11}gcm^{-3}$& $10^{11}gcm^{-3}$\\\hline
1 &  4.12E-02&    5.32E-04&  4.12E-02&    5.32E-04&  4.11E-02&    5.32E-04&  9.98E-01&    9.98E-01\\
3  & 2.36E+00&    7.05E-02&  2.36E+00&    7.06E-02&  2.35E+00&    7.03E-02&  2.33E+00&    7.03E-02\\
10 & 1.84E+00&    1.07E-01&  1.84E+00&    1.07E-01&  1.84E+00&    1.06E-01&  1.81E+00&    1.03E-01\\
30 & 8.15E+00&    4.79E-01&  8.17E+00&    4.80E-01&  8.15E+00&    4.79E-01&  7.96E+00&    4.67E-01\\
\hline
\end{tabular}
\end{table*}

The comparison with LSSM rates improves as one matches the results
from $^{54}$Fe to $^{55}$Fe and finally to $^{56}$Fe. In the case of
$^{56}$Fe the reported positron capture rates are suppressed by
around 2 orders of magnitude at T$_{9} =1$ (Table~\ref{ta10}). At higher
densities and temperatures the rates are in very good comparison.
The FFN rates are enhanced by roughly 4 orders of magnitude at low
temperatures and densities. The comparison improves as the density
and temperature of stellar core increases.

\section{Summary and conclusions}
The microscopic calculation of Gamow-Teller strength distributions
(GT$_{\pm}$) of three key isotopes of iron of astrophysical
importance was presented. The calculated strengths were in good
comparison to the measured strengths (for the case of $^{54,56}$Fe).
The pn-QRPA calculated total GT strengths were greater in magnitude
as compared to those using LSSM. The results also highlighted the
fact that the Brink's hypothesis and back resonances may not be a
good approximation to use in stellar calculation of weak rates.

The pn-QRPA model was also used to calculate the associated electron
and positron capture rates of these isotopes of iron with astrophysical importance.
Deformations of nuclei were taken into account and for the first
time the reported calculation took into consideration the
experimental deformations. The rates are calculated on an extensive
temperature-density grid point suitable for interpolation purposes
that might be required in collapse simulations. The electronic
versions of these files may be requested from the author.

During the oxygen and silicon core burning phase of massive stars
the pn-QRPA electron capture rates on $^{54}$Fe are around three
times bigger than those calculated by LSSM. The comparison with LSSM
gets better for proceeding pre-supernova and supernova phases of
stars. The pn-QRPA calculated electron capture rates on $^{55,56}$Fe
are in overall excellent agreement with the LSSM rates. The
calculated positron capture rates are generally smaller (by as much
as five orders of magnitude) at low temperatures and densities. The
calculation further discourages any noticeable contribution of
positron capture rates on iron isotopes for the physics of
core-collapse.

Due to the weak interaction processes (capture and decay rates) the
value of lepton-to-baryon ($Y_{e}$)  ratio for a massive star
changes from 1 (during hydrogen burning) to roughly 0.5 (at the
beginning of carbon burning)  and finally to around 0.42 just before
the collapse to a supernova explosion. The temporal variation of
$Y_{e}$ within the core of a massive star has a pivotal role to play
in the stellar evolution and a fine-tuning of this parameter at
various stages of presupernova evolution is the key to generate an
explosion. The electron capture tends to reduce this ratio whereas
the positron capture increases it. This paper reported on the
microscopic calculation of electron and positron capture rates on
iron isotopes and also highlighted the differences with previous key
calculations. What affect the calculated rates may have on the
simulation result is difficult to predict at this stage.
Core-collapse simulators are urged to check the affect of
incorporating pn-QRPA rates in their simulation codes for possible
interesting outcome.

\acknowledgments The author would like to acknowledge the kind
hospitality provided by the Abdus Salam ICTP, Trieste, where part of
this project was completed. The author wishes to acknowledge the
support of research grant provided by the Higher Education
Commission, Pakistan  through the HEC Project No. 20-1283.
\nocite{*}
\bibliographystyle{spr-mp-nameyear-cnd}
\bibliography{biblio-u1}

\begin{thebibliography}{}
\bibitem{Arn96} Arnet~D.: Supernovae and Nucleosynthesis. Princeton University
Press, Princeton, New Jersey. (1996)
\bibitem{Bet79} Bethe H.A., Brown G.E., Applegate J., Lattimer J.M.: Equation Of State In The Gravitational Collapse Of Stars. \nphysa
\textbf{324}, 487 (1979)
\bibitem{Iwa99} Iwamoto K., Brachwitz F., Nomoto K., Kishimoto N.,
Umeda H., Hix W.R., Thielemann F.-K.:  Nucleosynthesis in Chandrasekhar mass models for type IA supernovae and constraints on progenitor systems and burning-front propagation. \apjs
\textbf{125}, 439 (1999)
\bibitem{Roe93} R\"{o}nnqvist T., Cond\'{e} H., Olsson N., Ramstr\"{o}m E., Zorro R.,
Blomgren J., H\aa kansson A., Ringbom A., Tibell G., Jonsson O.,
Nilsson L., Renberg P.-U., van der Werf S.Y., Unkelbach W., Brady
F.P.: The $^{54,56}$Fe(n,p)$^{54,56}$Mn Reactions at E$_{n}$ = 97
MeV. \nphysa \textbf{563}, 225, (1993)
\bibitem{Vet89} Vetterli M.C., H\"{a}usser O., Abegg R., Alford W.P., Celler A., Frekers D., Helmer R.,
Henderson R., Hicks K.H., Jackson K.P., Jeppesen R.G., Miller C.A.,
Raywood K., Yen S.: Gamow-Teller Strength Deduced from Charge
Exchange Reactions on $^{54}$Fe at 300 MeV. \prc \textbf{40}, 559,
(1989)
\bibitem{Elk94} El-Kateb S., Jackson K.P., Alford W.P., Abegg R., Azuma
R.E., Brown B.A., Celler A., Frekers D., H\"{a}usser O., Helmer
R., Henderson R.S., Hicks K.H., Jeppesen R., King J.D., Raywood
K., Shute G.G., Spicer B.M., Trudel A., Vetterli M., Yen S.:
Spin-Isospin Strength Distributions for fp Shell Nuclei: Results for
the $^{55}$Mn(n,p), $^{56}$Fe(n,p), and $^{58}$Ni(n,p) Reactions at
198 MeV. \prc
\textbf{49}, 3128, (1994)
\bibitem{Rap83} Rapaport J., Taddeucci T., Welch T.P., Gaarde C., Larsen J.,
Horen D.J., Sugarbaker E., Koncz P., Foster C.C., Goodman C.D.,
Goulding C.A., Masterson T.: Excitation Of Giant Spin Isospin Multipole Vibrations In Fe-54,Fe-56 And Ni-58,Ni-60. \nphysa
 \textbf{410}, 371, (1983)
\bibitem{And90} Anderson B.D., Lebo C., Baldwin A.R., Chittrakarn T.,
Madey R., Watson J.W.: Gamow-Teller Strength in the
$^{54}$Fe(p,n)$^{54}$Co Reaction at 135 MeV, \prc
\textbf{41}, 1474 (1990)
\bibitem{Goo80} Goodman C.D., Goulding C.A., Greenfield M.B.,
Rapaport J., Bainum D.E., Foster C.C., Love W.G.,
Petrovich F.: Gamow-Teller Matrix-Elements From O-0 (P,N) Cross-Sections. \prl
\textbf{44}, 1755 (1980)
\bibitem{Ful82} Fuller G.M., Fowler W.A., Newman M.J.:
Stellar Weak-Interaction Rates for sd-Shell Nuclei. I. Nuclear
Matrix Element Systematics with Application to $^{26}$Al and
Selected Nuclei of Importance to the Supernova Problem. \apjs
\textbf{42}, 447 (1980);
Stellar Weak Interaction Rates
for Intermediate Mass Nuclei. III. Rate Tables for the Free Nucleons
and Nuclei with A = 21 to A = 60. \apjs
\textbf{ 48}, 279 (1982);
Stellar Weak Interaction
Rates for Intermediate Mass Nuclei. II. A = 21 to A = 60. \apj
\textbf{ 252}, 715 (1982);
Stellar Weak Interaction Rates for Intermediate Mass
Nuclei. IV. Interpolation Procedures for Rapidly Varying Lepton
Capture Rates Using Effective log (ft)- Values. \apj
\textbf{ 293}, 1 (1985)
\bibitem{Auf96} Aufderheide M.B., Bloom S. D., Mathews G. J., Resler
D. A.: Importance of (n,p) reactions for stellar beta decay rates.
\prc \textbf{53}, 3139 (1996)
\bibitem{Auf94} Aufderheide M.B., Fushiki I., Woosley S.E., Stanford
E., Hartmann D.H.: Search for Important Weak Interaction Nuclei
in Presupernova Evolution. \apjs
\textbf{91}, 389 (1994)
\bibitem{Lan00} Langanke K., Mart\'{i}nez-Pinedo G.: Shell-Model
Calculations of Stellar Weak Interaction Rates: II. Weak Rates for
Nuclei in the Mass Range A = 45-65 in Supernovae Environments. \nphysa
\textbf{673}, 481, (2000)
\bibitem{Nab04} Nabi J.-Un, Klapdor-Kleingrothaus H.V.: Microscopic
Calculations of Stellar Weak Interaction Rates and Energy Losses for
fp- and fpg-Shell Nuclei. At. Data Nucl. Data Tables
\textbf{ 88}, 237 (2004)
\bibitem{Joh92} Johnson C.W., Koonin S.E., Lang G.H., Ormand W.E.:
Monte-Carlo Methods For The Nuclear Shell-Model. \prl
\textbf{69}, 3157 (1992)
\bibitem{Hal67} Halbleib J. A., Sorensen R. A.: Gamow-Teller beta decay in heavy spherical nuclei and the unlike particle-hole rpa.
\nphysa \textbf{98}, 542 (1967)
\bibitem{Kru84}Krumlinde J., M\"{o}ller P.: Calculation of Gamow-Teller $\beta$-strength functions in the
rubidium region in the rpa approximation with Nilsson-model wave
functions. \nphysa \textbf{417}, 419 (1984)
\bibitem{Mut92} Muto K., Bender E., Oda T., Klapdor-Kleingrothaus H. V.: Proton-neutron quasiparticle RPA with separable Gamow-Teller forces.  Z.
Phys. A \textbf{341}, 407 (1992)
\bibitem{Nab99} Nabi J.-Un, Klapdor-Kleingrothaus H.V.: Weak
Interaction Rates of sd-Shell Nuclei in Stellar Environments
Calculated in the Proton-Neutron Quasiparticle Random-Phase
Approximation. At. Data Nucl. Data Tables
\textbf{ 71}, 149 (1999)
\bibitem{Nab99a} Nabi J.-Un, Klapdor-Kleingrothaus H.V.: Microscopic
Calculations of Weak Interaction Rates of Nuclei in Stellar
Environment for A = 18 to 100. Eur. Phys. J. A
\textbf{5}, 337 (1999)
\bibitem{Nab05} Nabi J.-Un, Rahman M.-Ur.:  Gamow-Teller Strength
Distributions and Electron Capture Rates for $^{55}$Co and
$^{56}$Ni. Phys. Lett.
\textbf{B612}, 190 (2005)
\bibitem{Nab07} Nabi J.-Un, Rahman M.-Ur.: Gamow-Teller
Transitions from $^{24}$Mg and Their Impact on the Electron Capture
Rates in the O+Ne+Mg Cores of Stars. \prc
\textbf{75}, 035803 (2007)
\bibitem{Nab07a} Nabi J.-Un, Sajjad M., Rahman M.-Ur.: Electron
Capture Rates on Titanium Isotopes in Stellar Matter. Acta Phys. Pol. B
\textbf{38}, 3203 (2007)
\bibitem{Nab07b} Nabi J.-Un, Sajjad M.: Comparative Study of Gamow-Teller
Strength Distributions in the Odd-odd Nucleus $^{50}$V and its
Impact on Electron Capture Rates in Astrophysical Environments. \prc
\textbf{76}, 055803 (2007)
\bibitem{Nab08} Nabi J.-Un, Rahman M.-Ur, Sajjad M.: Gamow-Teller
(GT$_{\pm}$) Strength Distributions of $^{56}$Ni for Ground and
Excited States. Acta Phys. Pol. B
\textbf{39}, 651 (2008)
\bibitem{Nab08a} Nabi J.-Un, Sajjad M.: Expanded Calculations
of Proton-Neutron Quasiparticle Random Phase Approximation (pn-QRPA)
Electron Capture Rates on $^{55}$Co for Presupernova and Supernova
Physics. Can. J. Phys.
\textbf{86}, 819 (2008)
\bibitem{Nab08b} Nabi J.-Un, Sajjad M.: Neutrino Energy Loss Rates
and Positron Capture Rates on $^{55}$Co for Presupernova and
Supernova Physics. \prc
\textbf{77}, 055802 (2008)
\bibitem{Nab09}Nabi J.-Un.: Weak-Interaction-Mediated Rates on
Iron Isotopes for Presupernova Evolution of Massive Stars. Eur. Phys. J. A
\textbf{40}, 223 (2009)
\bibitem{Nab10} Nabi J.-Un.: Expanded calculation of neutrino cooling rates due to $^{56}$Ni in stellar matter. \physscr
\textbf{81}, 025901 (2010)
\bibitem{Nab10a} Nabi J.-Un.:Neutrino and antineutrino energy loss rates in massive stars due to isotopes of titanium.
 Int. J. Mod. Phys. E \textbf{19}, 1 (2010)
\bibitem{Heg01} Heger A., Woosley S.E., Mart\'{i}nez-Pinedo G.,
Langanke K.:  Presupernova Evolution with Improved Rates for Weak
Interactions. \apj
\textbf{ 560}, 307 (2001)
\bibitem{Sta90} Staudt A., Bender E., Muto K.,
Klapdor-Kleingrothaus H.V.: Second-Generation Microscopic
Predictions of Beta-Decay Half-lives of Neutron-Rich Nuclei. At. Data Nucl. Data Tables
\textbf{ 44}, 79 (1990)
\bibitem{Hir93} Hirsch M., Staudt A., Muto K., Klapdor-Kleingrothaus H.V.:
Microscopic Predictions of $\beta^{+}$/EC-Decay Half-Lives. At. Data Nucl. Data Tables
\textbf{ 53}, 165 (1993)
\bibitem{Nil55} Nilsson S.G.: Binding States of Individual Nucleons in
Strongly Deformed Nuclei. Mat. Fys. Medd. Dan. Vid. Selsk
\textbf{ 29}, 16 (1955)
\bibitem{Hir91} Hirsch M., Staudt A., Muto K., Klapdor-Kleingrothaus H.V.:
Microscopic Calculation of $\beta^{+}$/EC-Decay Half-Lives with
Atomic Numbers Z = 10-30. \nphysa
\textbf{535}, 62 (1991)
\bibitem{Ste04}Stetcu I., Johnson C.W.: Gamow-Teller Transitions
and Deformation in the Proton-Neutron Random Phase Approximation. \prc
\textbf{69}, 024311 (2004)
\bibitem{Ram87} Raman S., Malarkey C.H., Milner W.T., Nestor, Jr. C.W., Stelson
P.H.: Transition Probability, B(E2)$\uparrow$, from the Ground to
the First-Excited 2$^{+}$ State of Even-Even Nuclides. At. Data Nucl. Data Tables
\textbf{36}, 1 (1987)
\bibitem{Moe81}M\"{o}ller P., Nix J.R.: Atomic Masses and Nuclear
Ground-State Deformations Calculated with a New
Macroscopic-Microscopic Model. At. Data Nucl. Data Tables
\textbf{26}, 165 (1981)
\bibitem{Aud03}Audi G., Wapstra A.H., Thibault C.: The AME2003
Atomic Mass Evaluation (II). Tables, Graphs and References. \nphysa
\textbf{729}, 337 (2003)
\bibitem{Yos88} Yost G.P., Barnett R.M., Hinchliffe I.,  {\it et al.}
(Particle Data Group), Review of Particle Properties, Phys. Lett.
\textbf{ B204}, 1 (1988)
\bibitem{Mut89} Muto K., Bender E., Klapdor H.V.: Proton-Neutron
Quasiparticle RPA and Charge-Changing Transitions. Z. Phys. A
\textbf{333}, 125 (1989)
\bibitem{Ost92} F. Osterfeld.: Nuclear-Spin And Isospin Excitations.
Rev. Mod. Phys. \textbf{64}, 491 (1992)
\bibitem{Gaa83} Gaarde C.: Gamow-Teller and M1 Resonances. \nphysa
\textbf{396}, 127c (1983)
\bibitem{Rod06} Rodin V., Faessler A., Simkovic F., Vogel P.:
Uncertainties in the 0-Decay Nuclear Matrix Elements, Czech. J. Phys
\textbf{ 56}, 495 (2006)
\end{thebibliography}

\end{document}